\documentclass[pdftex,twocolumn,epjc3]{svjour3}          

\RequirePackage[T1]{fontenc}

\usepackage{subfigure,dcolumn}
\usepackage[T2A,T1]{fontenc}
\usepackage[russian,english]{babel}
\usepackage{amsmath}
\usepackage{verbatim}
\usepackage{listings}
\usepackage{makecell}
\usepackage{mathrsfs}
\usepackage{amsbsy}
\usepackage{ulem}
\usepackage{multirow}
\usepackage{booktabs} 
\usepackage{float}
\usepackage{color}
\usepackage{soul}
\usepackage{xcolor}
\smartqed  

\RequirePackage{graphicx}
\RequirePackage{mathptmx}      
\RequirePackage{flushend}
\RequirePackage[numbers,sort&compress]{natbib}
\RequirePackage[colorlinks,citecolor=blue,urlcolor=blue,linkcolor=blue]{hyperref}

\journalname{Eur. Phys. J. C}

\usepackage{lineno}

\begin{document}

\title{Reconstruction of the Effective Energy-deposition Vertex of Muon Showers using PMT Waveform in a Large-scale Liquid Scintillator Detector}


\author{Junwei Zhang\thanksref{e1,addr1,addr2}, Yongpeng Zhang\thanksref{e2,addr2}, Yongbo Huang\thanksref{e3,addr1}, Jilei Xu\thanksref{e4,addr2},  Junyou Chen\thanksref{e5,addr1,addr2} \and Yi Wang\thanksref{e6,addr2} 
}

\thankstext{e1}{e-mail: zhangjunwei@st.gxu.edu.cn}
\thankstext{e2}{e-mail: ypzhang1991@ihep.ac.cn (corresponding author)}
\thankstext{e3}{e-mail: huangyb@gxu.edu.cn (corresponding author)}
\thankstext{e4}{e-mail: xujl@ihep.ac.cn (corresponding author)}
\thankstext{e5}{e-mail: chenjy2022@st.gxu.edu.cn}
\thankstext{e6}{e-mail: wangyi90@ihep.ac.cn}


\institute{Guangxi Key Laboratory for Relativistic Astrophysics, School of Physical Science and Technology, Guangxi University, Nanning 530004, China \label{addr1}
           \and
           Institute of High Energy Physics, Beijing 100049, China. \label{addr2}
}

\date{Received: date / Accepted: date}

\maketitle

\begin{abstract}
Cosmogenic muon-induced radioactive isotopes pose a significant background source for deep-underground low-background experiments. Although rock overburdens at underground sites substantially attenuate the cosmogenic muon flux, residual muon-induced backgrounds still require active suppression. For future multi-kiloton liquid scintillator (LS) detectors, such as the Jiangmen Underground Neutrino Observatory (JUNO), shower muons contribute to more than 88\% of all muon-induced isotopes. Consequently, precise reconstruction of shower vertices is essential for implementing localized spatial vetoes. We propose a novel waveform-based method to reconstruct the shower vertex, defined as the energy-deposition centroid in radius of 2~m. By subtracting the track contributions from non-shower muons in the recorded waveforms, the isolated shower component is extracted. Subsequently, using a photon propagation model and an iterative optimization algorithm, the shower vertex positions are reconstructed. Simulations show that for 68\% of events, the single shower vertex resolution is better than 0.16~m, 0.15~m, and 0.26~m along X, Y, and Z respectively. Furthermore, the reconstruction efficiency exceeds 96\% when requiring the distance between the reconstructed and true vertices to be less than 3.0 m. This method provides a critical technical foundation for muon-induced background suppression in JUNO and other large-scale LS detectors.

\end{abstract}

\section{Introduction}
\label{sec:intro} 

Cosmogenic muon-induced spallation produces radioactive isotopes, which constitute significant background sources in low-background experiments. Although detectors are generally shielded by being located in underground experimental facilities, where substantial rock overburden effectively attenuates the majority of cosmogenic muons, residual muon-induced backgrounds still require additional suppression. Consequently, muon tagging techniques and corresponding veto strategies are essential. This approach is widely adopted in underground experiments (e.g., KamLAND~\cite{KamLAND-2009zwo}, Borexino~\cite{Borexino-2013cke}, Daya Bay~\cite{DayaBay-2014cmr}, Double Chooz~\cite{DoubleChooz-2014qeg}, SNO+~\cite{Hackenburg-2017syz}).

\begin{figure}[!htb]
\centering
\includegraphics[width=0.98\linewidth]{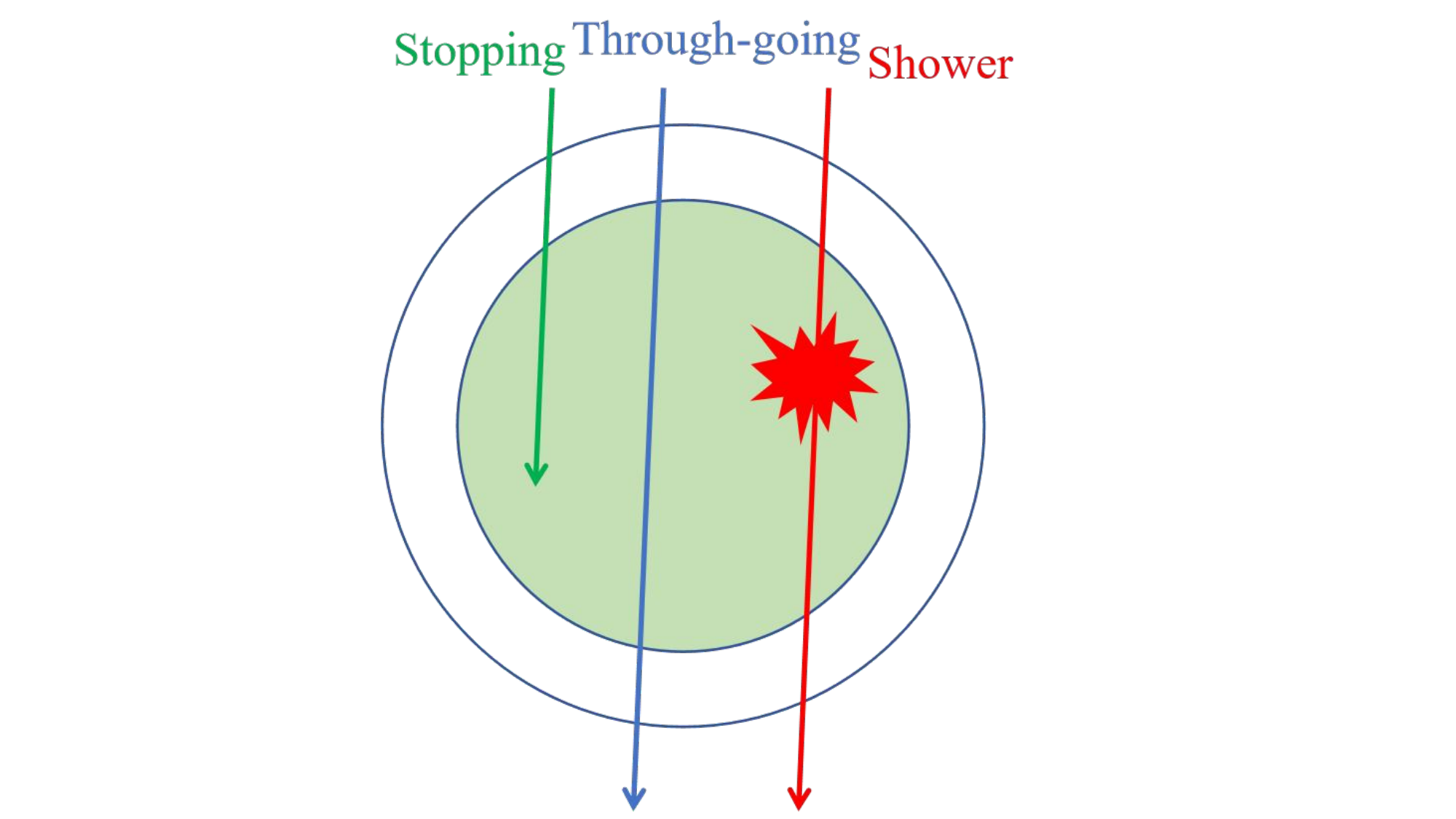}
\caption{Schematic of a spherical target detector and different kinds of muon events labelled with shower muon, through-going muon, and stopping muon}
\label{fig:muon_types}
\end{figure}

The muon veto strategy typically exploits the spatio-temporal correlation between a muon track and its induced isotopes. By vetoing a volume surrounding the muon track for a specific time window, these isotopes can be effectively rejected. Consequently, the development of efficient muon tagging techniques is a prerequisite for implementing the muon veto. Muon events can be categorized according to their topology in the detector. Primary classifications include through-going muons (non-shower muons), shower muons (most of them can go through detector), and stopping muons (almost no shower), as illustrated in Fig.~\ref{fig:muon_types}. Further distinctions, such as clipping muons and muon bundles, arise when considering factors like the muon's path relative to the detector and its multiplicity. Comprehensive classification schemes are detailed in~\cite{Yang-2022din}.

In prior experiments utilizing detectors with size under 10~m, the constrained size limited the muon type predominantly to through-going muons, characterized by low muon rates (typically below 1 Hz). Consequently, implementing a full-detector veto for a defined time window preserved high live time and detection efficiency. For example, the attenuated muon fluxes in the three experimental halls of the Daya Bay experiment are 1.16, 0.86, and 0.054 Hz/m$^{2}$, respectively~\cite{DayaBay-2014cmr}. The live time of the antineutrino detectors (size $\sim$~4~m) in these halls is reduced following the application of the muon veto, resulting in corresponding efficiencies of approximately 82.4\%, 85.7\%, and 98.2\%~\cite{DayaBay-2016ggj}. However, this strategy becomes highly inefficient for kiloton-scale LS detectors. Instead, reconstructing the muon trajectory using deposited charge and timing information, followed by vetoing a cylindrical volume centered on the reconstructed track, proved significantly more effective. At the kiloton scale, single through-going muons remain the dominant type, and their distinct linear track signatures facilitate relatively straightforward reconstruction. The deployment of 10-kiloton class detectors, however, reveals non-negligible occurrences of muon bundles-multiple muons traversing the detector simultaneously. Furthermore, a larger fraction of incident muons induce energetic showers (hadronic or electromagnetic), classified as shower muons. These evolving characteristics create stringent demands on muon reconstruction precision and the effective rejection of muon-induced isotope backgrounds in low-background experiments. As highlighted in~\cite{Grassi-2014hxa}, the full-detector veto approach is inadequate under these conditions. This necessitates the development of advanced muon reconstruction and classification techniques capable of handling diverse muon event topologies, including single muons, muon bundles, and shower muons.

The Jiangmen Underground Neutrino Observatory (JUNO), a LS-based experiment designed to determine the neutrino mass ordering, has been completed construction and has begun acquiring data~\cite{JUNO-2025fpc,JUNO-2025gmd}. It is now operating the world's largest liquid scintillator (LS) detector, with a target mass of 20 kilotons. About 17600 20-inch photomultiplier tubes (PMTs) and 25600 3-inch PMTs are equipped as photosensors. Additional detector details are provided in~\cite{JUNO-PPNP, JUNO-PMT}. A rock overburden of approximately 650 m (1,800 m.w.e.) above the JUNO central detector (CD) reduces the muon flux by about four orders of magnitude to muon rate of approximately 0.004 Hz/m$^{2}$ and an average muon energy of around 207 GeV at JUNO ~\cite{JUNO-PPNP}. However, the radioactive isotopes produced by muons remain significant contributors to backgrounds in both reactor neutrino and solar neutrino detection. These isotopes are including $^{9}$Li, $^{8}$He, $^{11}$C and other unstable nuclides. Among these isotopes with interested energy regions in JUNO, $\beta$-n cascade decay from $^{9}$Li, $^{8}$He will contribute major background in inverse beta decay detection channel and the decay signal with from other isotopes will form accidental coincidence background. Therefore, it is essential to further implement muon tagging technology and corresponding muon veto strategies to mitigate the remaining muon-induced backgrounds. For the JUNO experiment, extensive studies have been conducted on the track reconstruction of single through-going muons~\cite{Genster-2018caz,Zhang-2018kag,Liu-2021okf} and double through-going muons~\cite{Yang-2022din}. However, there remains a scarcity of detailed reports regarding the reconstruction of shower muons in JUNO.

The references~\cite{Li-2014sea, Li-2015kpa, Li-2015lxa} carried out extensive analyses of shower muon physics in large water Cherenkov detectors, demonstrating that the majority of muon-induced isotopes are generated by secondary particles produced in muon-induced showers, rather than by the incident cosmogenic muons themselves. These isotopes are predominantly located within several meters of the energy-deposition centroid. This spatial correlation suggests that reconstructing the weighted center of the shower energy deposition may offer a more effective strategy for background rejection, with minimal impact on detector dead time. In Section~\ref{sec:Characteristics-of-shower}, detailed simulation results for shower muons in JUNO will be presented, where the spatial relationship between muon-induced isotopes and the shower can be examined, providing the foundation for this definition. Based on these studies, Super-K developed advanced background rejection techniques that substantially improved physics measurement precision~\cite{Super-Kamiokande-2021snn}. Similarly, the KamLAND-Zen experiment implemented shower muon reconstruction in its large-scale LS detector~\cite{KamLAND-Zen-2023spw}. Compared to technologies like the Liquid Argon Time Projection Chamber (LArTPC) used in the DUNE experiment, which has great tracking capability and precise measurement on energy deposits~\cite{DUNE-2020mra}, LS detectors face challenges due to isotropic lighting, making it difficult to detect the profile of a shower. Consequently, reconstructing shower vertices in LS detectors is particularly challenging. 

This paper presents the first comprehensive study of shower muon reconstruction in a 20-kton LS detector, using the JUNO experiment as a benchmark. Our methodology provides critical technical support for muon-induced background rejection in JUNO and future large-volume, low-background experiments. The paper is structured as follows: Section~\ref{sec:showerCharacter} characterizes shower muon simulations, event topologies, and isotope yields by muon species, and analyzes the PMT waveform signatures that form the foundation of our reconstruction technique. Section~\ref{sec:method} details our reconstruction algorithm. Reconstruction performance is evaluated in Section~\ref{sec:Performance}, with results and discussion presented in Section~\ref{sec:conclusion}.

\section{Characteristics of shower muon in LS}
\label{sec:showerCharacter}

\subsection{Simulation description}
\label{sec:simulation}

\begin{figure}[!htb]
        \centering
        \includegraphics[width=0.7\linewidth]{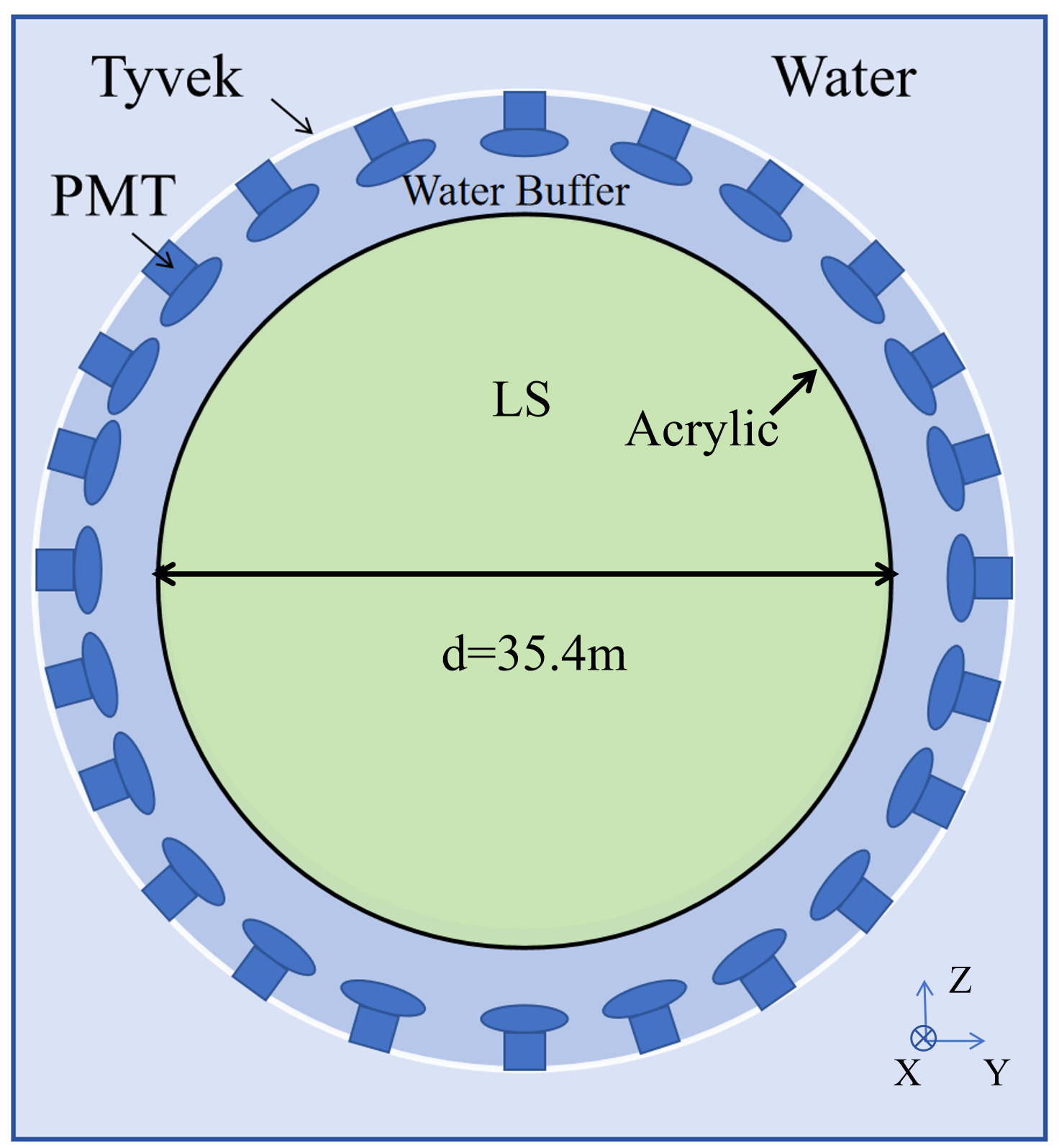}
        \caption{A schematic view of the detector in the simulation. The Z-axis is the vertical central axis of the detector and the origin of the coordinate system is the center of the detector.}
        \label{fig:JUNO_schematic}
    \end{figure} 

\begin{figure*}[!htb]
        \centering
        \subfigure[]{
            \includegraphics[width=0.4\hsize]{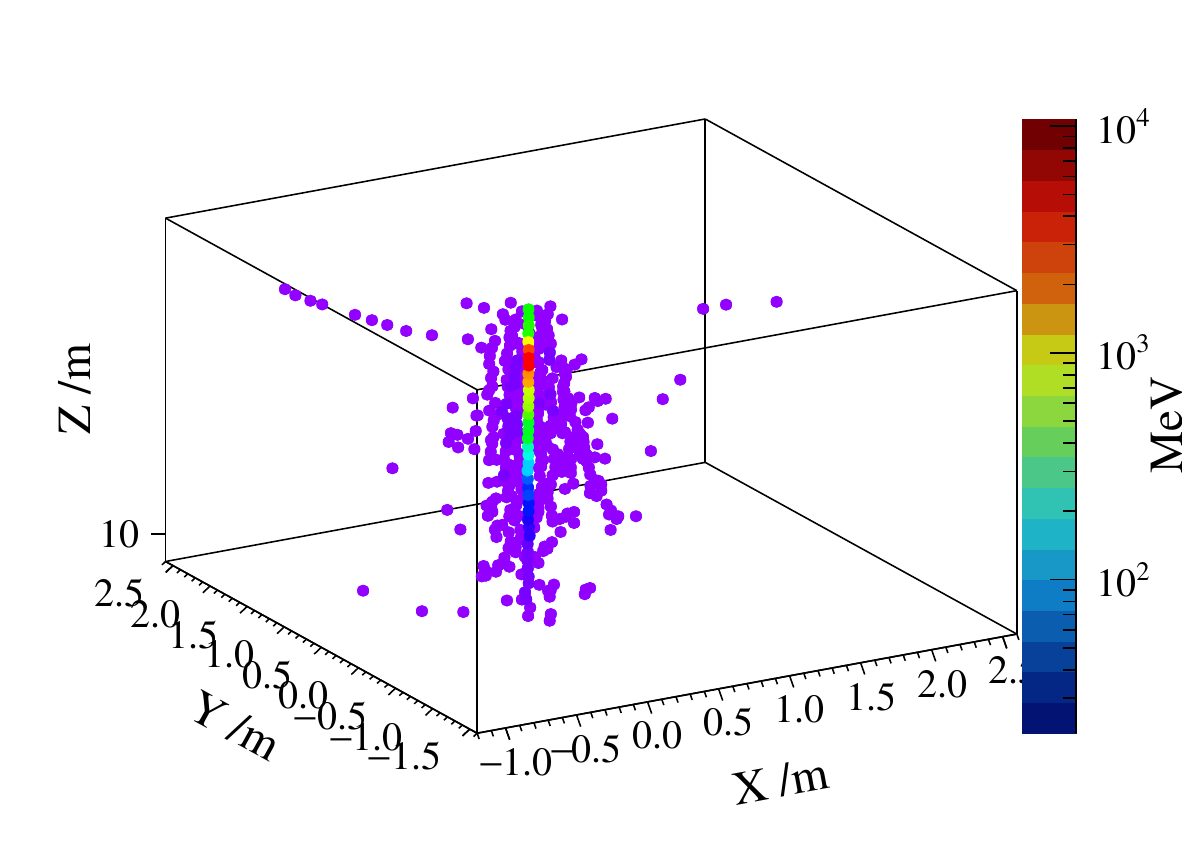}
        	\label{fig:longitudinalextension}
        }
        \subfigure[]{
            \includegraphics[width=0.4\hsize]{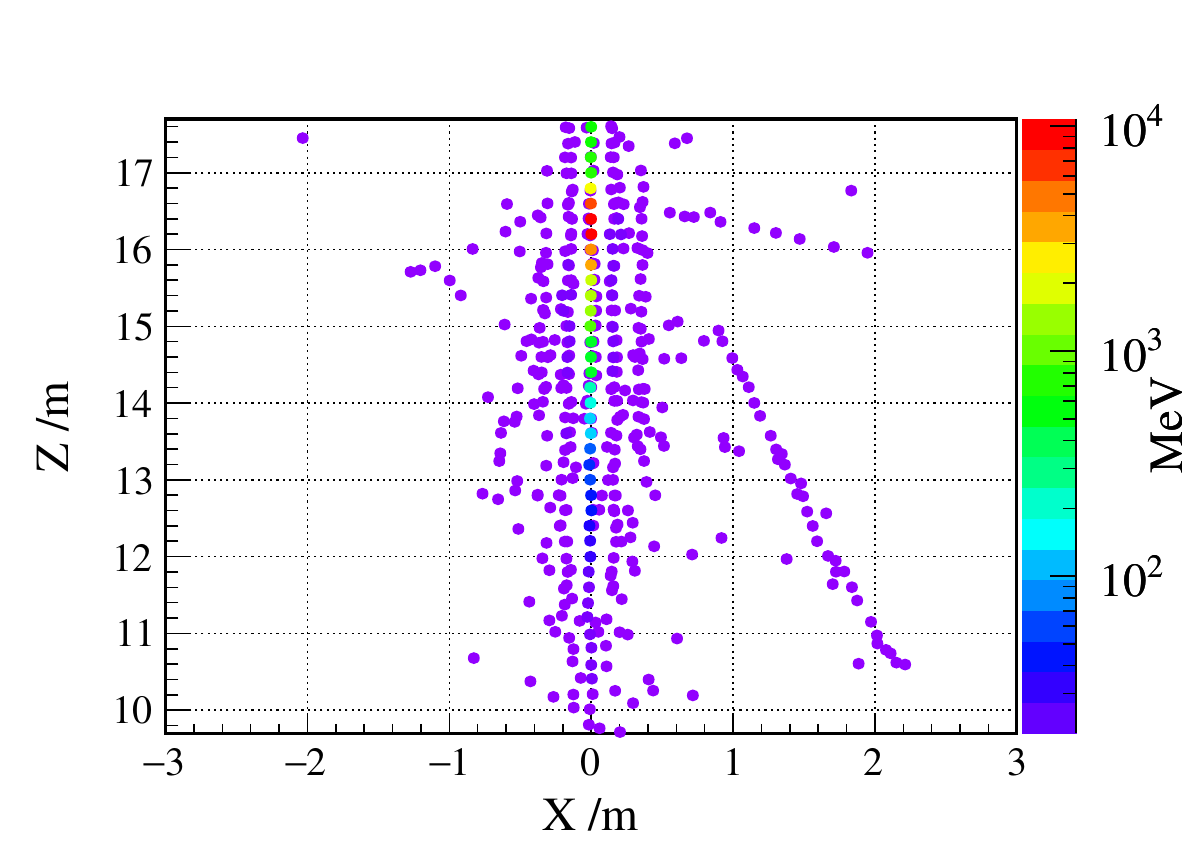}
        	\label{fig:longitudinalextension}
        }
        \subfigure[]{
            \includegraphics[width=0.4\hsize]{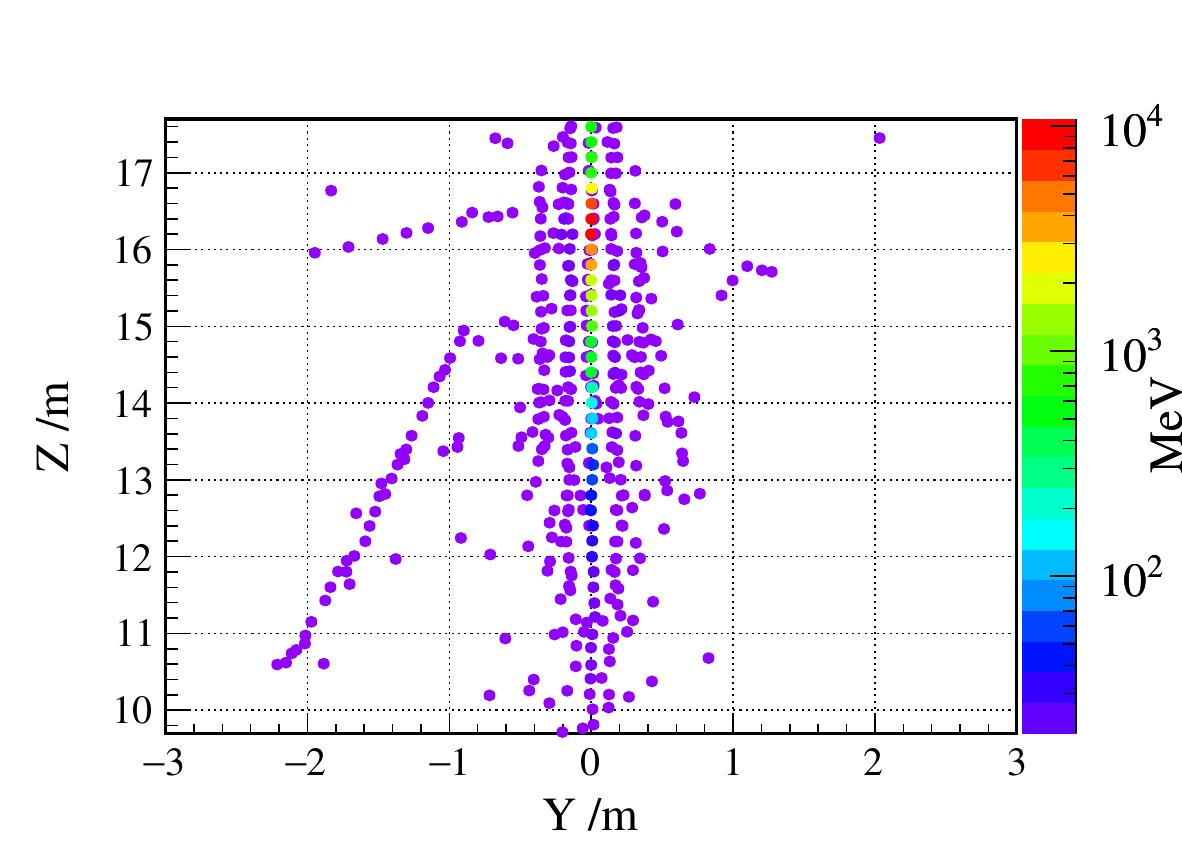}
        	\label{fig:voxelDistanceToCenter}
         }
        \subfigure[]{
            \includegraphics[width=0.4\hsize]{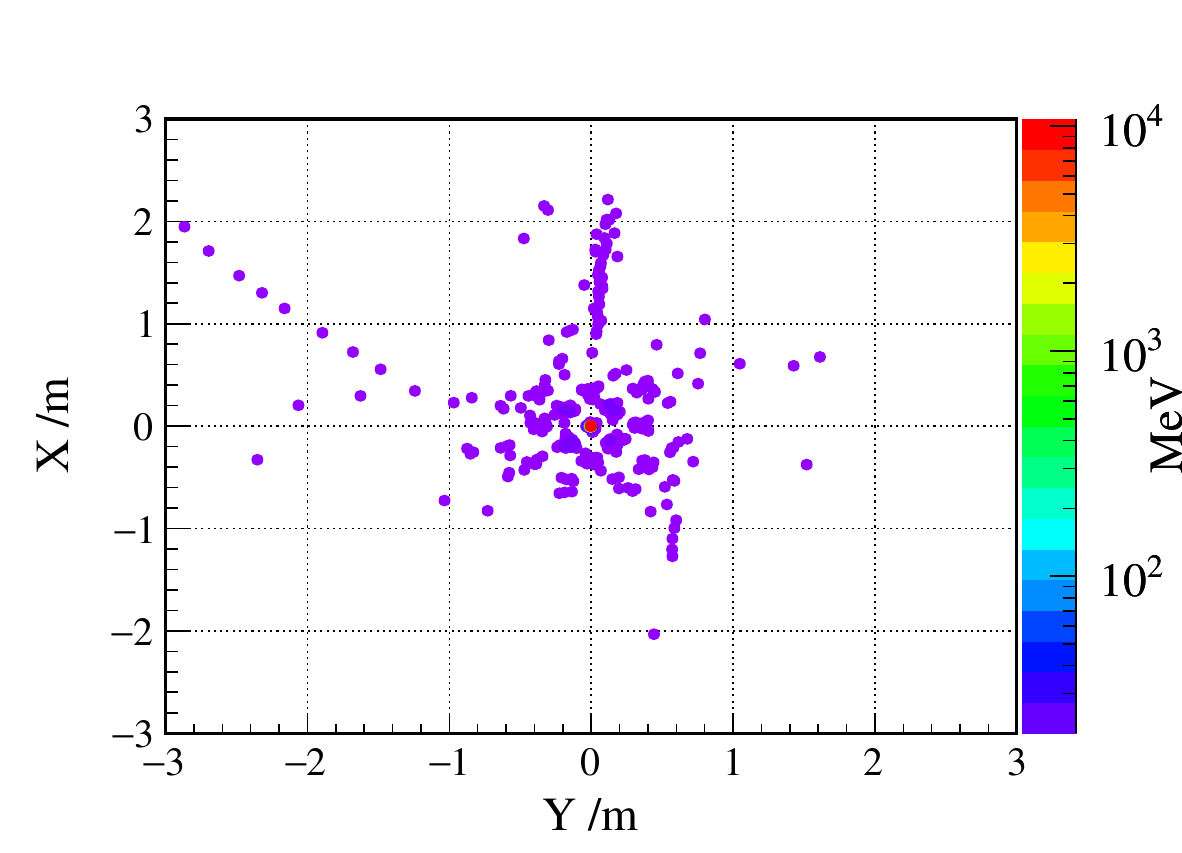}
        	\label{fig:shower_xz_plane}
        }
        \caption{An example of the energy deposition expansion in three-dimensional space for a shower muon event ($\sim$ 220~GeV), where the energy deposition includes both track components and shower components. For clarity, the two dimension plots, XZ, YZ, YX are also shown.} 
    	\label{fig:shower_muon_example_3-2Dplots}
\end{figure*}

A simulation software framework based on Geant4~\cite{GEANT4-2002zbu,Allison-2006ve,Allison-2016lfl, juno-sim} was developed for detector simulation, using the JUNO-like detector geometry as a benchmark to investigate shower characteristics of muons in the liquid scintillator. The simulated detector layout is illustrated in Fig.~\ref{fig:JUNO_schematic}. Muon events were generated based on Gaisser formula~\cite{gaisser} at the ground surface and transported in the mountain profile by MUSIC software~\cite{ANTONIOLI1997357} to the underground 650~m position. The underground muon events were imported into the Geant4 simulation, and then shower muons (defined later) were selected for further analysis. To comprehensively record energy deposits from muons and their secondary particles, the liquid scintillator detector was segmented into voxels of size 30~cm $\times$ 30~cm $\times$ 30~cm.
Consequently, the shower energy, along with its deposition location and spatial distribution range, can be effectively visualized. Fig.~\ref{fig:shower_muon_example_3-2Dplots} shows an example of the space extension of a shower muon event going through LS. Each point is for the center position of the voxel and the colour is for the total deposited energy larger than 20~MeV in a voxel. The two-dimensional projections clearly show the lateral extension around the muon track. Along the longitudinal direction, the shower extends further forward.

\begin{figure*}[!htb]
        \centering
        \subfigure[]{
            \includegraphics[width=0.3\hsize]{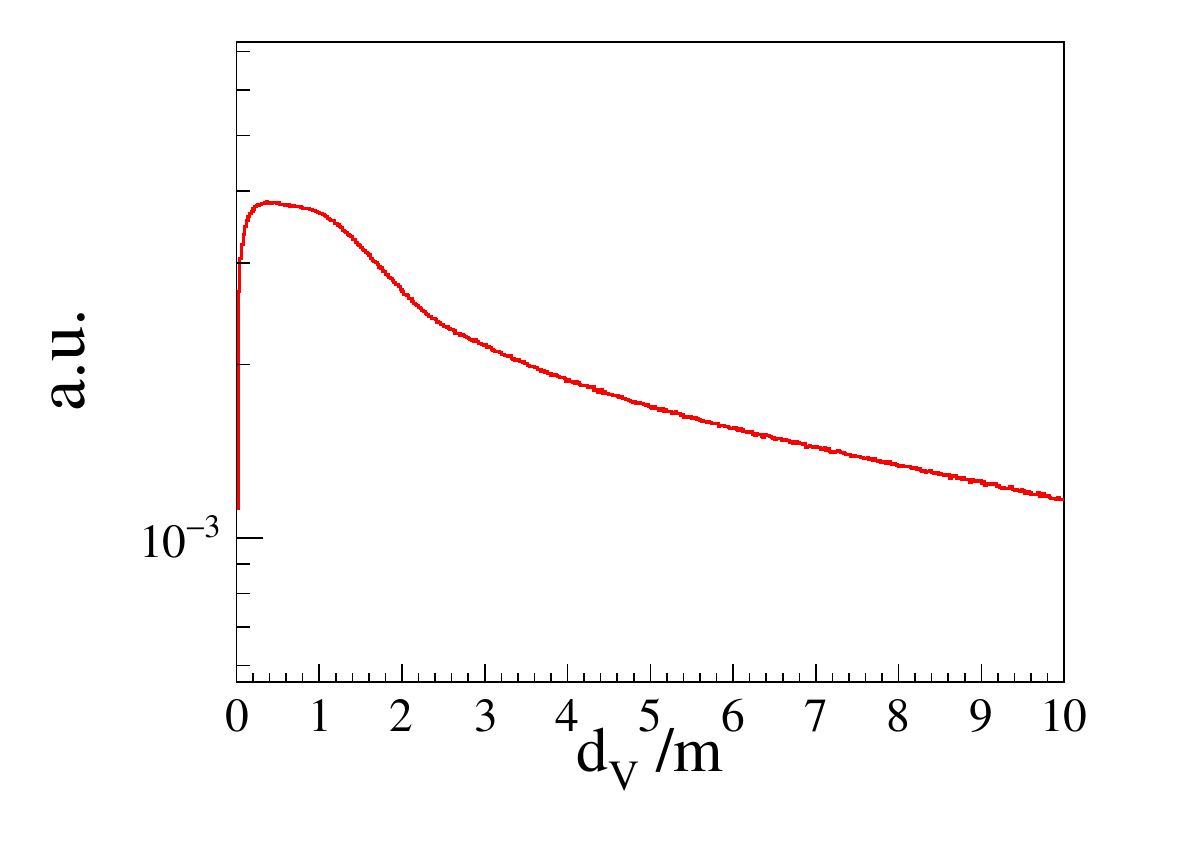}
        	\label{fig:voxelDistanceToCenter}
        }
        \subfigure[]{
            \includegraphics[width=0.3\hsize]{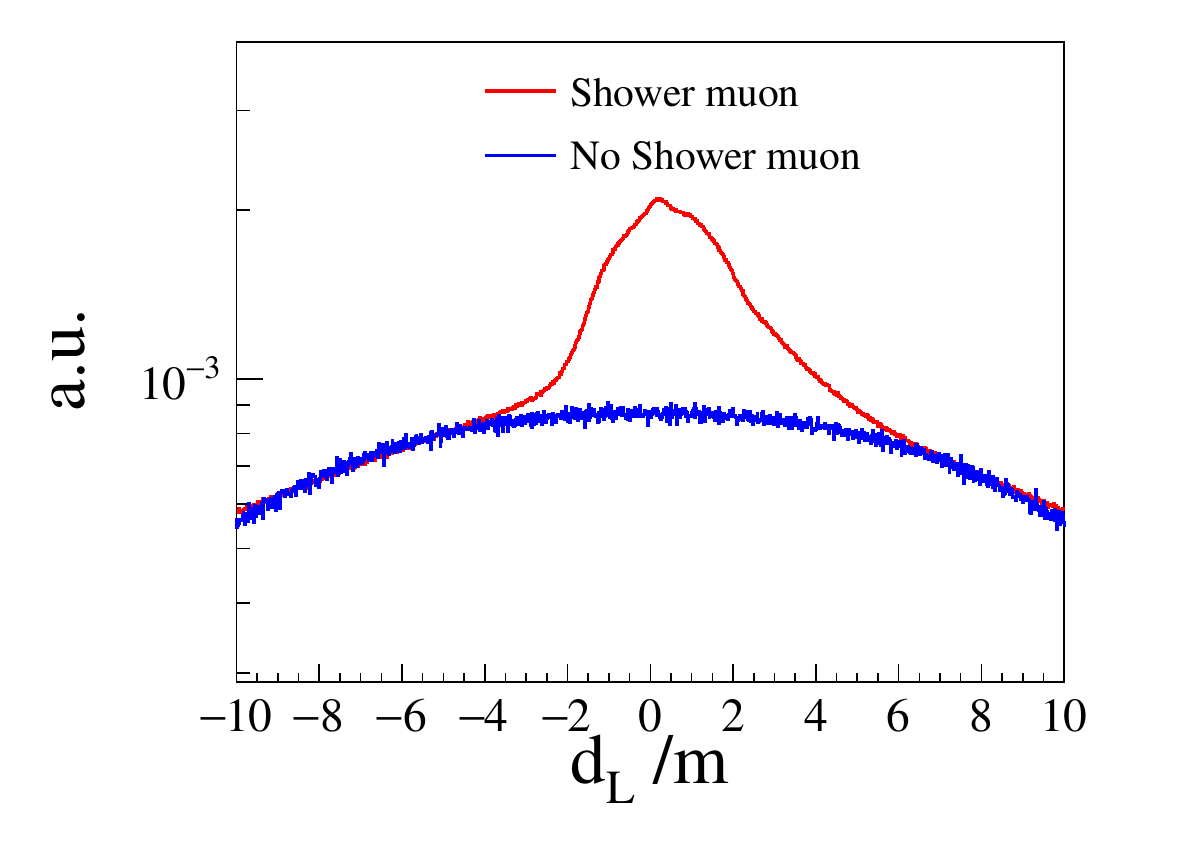}
        	\label{fig:longitudinalextension}
         }
        \subfigure[]{
            \includegraphics[width=0.3\hsize]{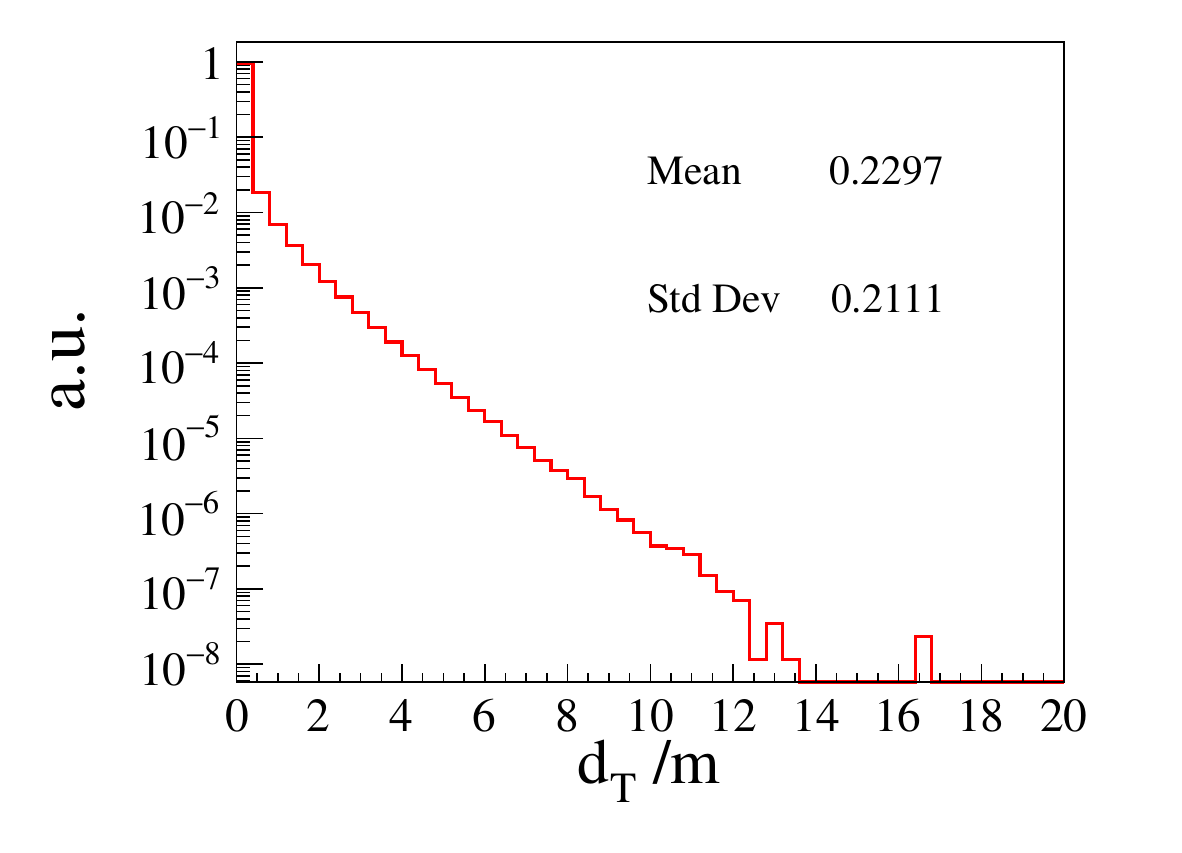}
        	\label{fig:voxelDistanceToTrack}
        }
        \caption{The shower muon energy deposit profiles based on simulation information. (a) The distance (d$_{V}$) distribution of the energy deposited voxels relative to the shower vertex. (b) The longitudinal distance (d$_{L}$) distribution of the energy deposited voxels relative to the muon track. The positive (negative) values are the voxels project onto the forward (backward) place of the shower vertex in the muon track. (c) The transverse distance (d$_{T}$) distribution from energy deposited voxels to muon track} 
    	\label{fig:muonShowerProfile}
\end{figure*}

According to our study, when a muon deposits a significant amount of energy at a given interaction vertex, the energy deposited in surrounding small voxels follows a declining trend with increasing distance. To give a clear explaination, the shower vertex, shower muon, and shower energy are defined as the following: 
\begin{itemize}
    \item \textbf{Shower vertex}: The deposit energy-weighted center of voxels within a radius of 2~m. This range can cover most of the deposited energy of the shower~\cite{Nairat-2025sju}. If the distance of two shower vertexes is larger than 4~m, this muon has two shower vertexes.
    \item \textbf{Shower muon}: The muon that accumulated deposited energy larger than 1.8~GeV in a sphere with radius 2~m.
    \item \textbf{Shower energy}: The accumulated deposited energy subtract the muon track minimum ionization energy in a sphere with radius 2~m.
\end{itemize}

Figure~\ref{fig:muonShowerProfile} presents the energy deposit profiles of shower muons. Figure~\ref{fig:voxelDistanceToCenter} presents the distribution of distances (d$_{V}$) from each energy-depositing voxel to the reconstructed shower vertex. This reflects the radial spread of the shower around its energy-weighted center~\cite{Zhang-2024okq}, with the inflection point near 2~m indicating the typical spatial extent of the electromagnetic and hadronic components. Figure~\ref{fig:longitudinalextension} shows the longitudinal distribution (d$_{L}$) of energy deposition relative to the muon track direction. Values are calculated along the track, with positive (negative) ($d_{L}$) corresponding to positions ahead of (behind) the shower vertex. The asymmetric shape and inflection point illustrate the forward-peaking nature of shower development due to the high-energy muon's direction. Figure~\ref{fig:voxelDistanceToTrack} displays the transverse profile (d$_{T}$), representing the perpendicular distance from deposited energy voxels to the muon track. The observed exponential decay is consistent with expectations from particle cascade theories and provides useful insight into shower development and supports reconstruction efforts.

Based on the distributions shown in Fig. ~\ref{fig:voxelDistanceToCenter} and Fig.~\ref{fig:longitudinalextension}, a sphere can work as a simple first-order approximation of the shape of a shower. The inflection point near 2~m  motivates the radius for identifying shower development. A muon is classified as a shower muon if the total energy deposited within this sphere exceeds 1.8~GeV—a threshold selected to include the muon's minimum ionization energy (roughly 2~MeV/cm and ~800~MeV over 4~m) plus an additional 1~GeV from shower secondaries. To prevent the merging of nearby showers, a minimum separation of 4 m is required between any two shower vertices.

Additionally, an electronics simulation was implemented to facilitate the reconstruction of shower vertices. The simulation models the response of the PMTs by superimposing the single photoelectron (SPE) waveforms. PMT parameters — including SPE charge smearing, transit time spread (TTS), dark noise, and afterpulse — were incorporated into the simulation based on data obtained from PMT testing~\cite{JUNO-2022hlz}. Furthermore, electronic effects such as white noise and dynamic range were taken into account. White noise was generated randomly according to a Gaussian distribution, with a $\sigma$ value approximately one-tenth of the amplitude of the PMT SPE signal~\cite{JUNO-2022hlz,JUNO-2025fpc}. Following these operations, the waveforms of muon events can be generated for subsequent analysis.

\subsection{Characteristics of shower muon and isotopes yield in LS}
\label{sec:Characteristics-of-shower}

After defining the shower muon and the shower vertex, we can analyze the characteristics of the shower muon using the information recorded in the simulation. 

\begin{figure}[!htb]
\centering
\includegraphics[width=0.9\linewidth]{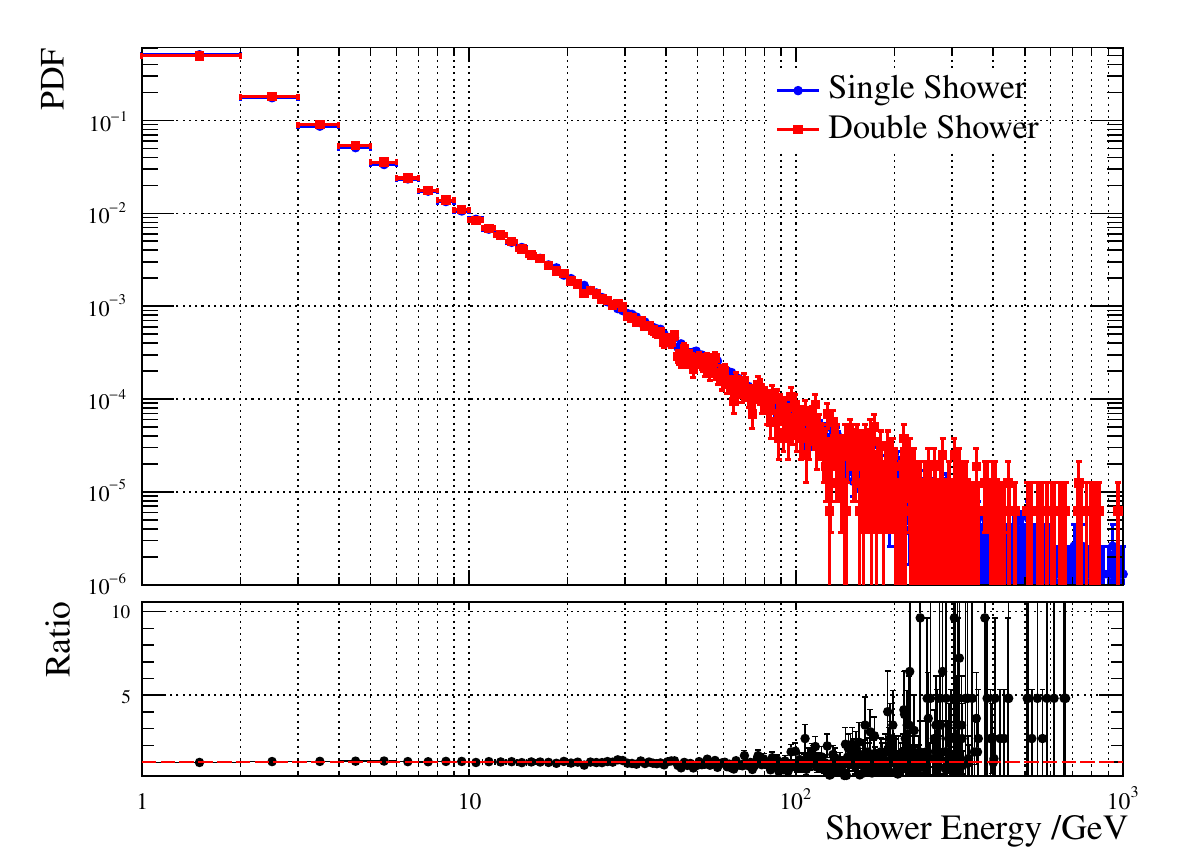}
\caption{The above panel shows the shower energy spectrum for each shower in the LS. The bottom panel shows the ratio of individual showers from double shower events to single shower events}
\label{fig:singleMultiShower}
\end{figure}

For shower muon events, the average energy deposition in the LS exceeds that of non-shower muons. In the spherical LS volume with 35.4~m diameter, the total deposited energy for minimally ionizing muons is approximately 7~GeV, whereas shower muons deposit $>~10$~GeV on average. Fig.~\ref{fig:singleMultiShower} presents the energy spectrum per shower within a sphere with radius 2~m. Single-shower events contain one distinct vertex along the muon trajectory through the LS, while double-shower events feature two spatially separated vertices whose energies are recorded independently. The energy distribution of individual showers in double-shower events closely matches that of single showers, indicating stochastic shower generation where double showers effectively constitute two independent single showers. This analysis focuses primarily on single-shower events, with supplementary discussion of double-shower scenarios.

Based on simulations using a LS detector with a diameter of 35.4~m, the average shower multiplicity per muon (number of muon shower) is approximately 1.6 when an energy threshold above 1.8~GeV is applied within a 2~m distance. Under these conditions, the number of muon showers can reach up to 4 in the LS. The average distance between two shower vertices is found to be about 13~m.

\begin{table}[h]
\centering
\caption{\label{tab:iso} The proportion of isotopes produced by different shower energies of single muon events and by non-shower muon events}
\begin{tabular*}{\columnwidth}{l@{\extracolsep{\fill}}lcc}
\hline
\multicolumn{2}{c}{Muon event type} & Muon ratio & Isotope ratio \\
\hline
\multirow{6}{*}{Shower muon} & All & 19.8\% & 88.2\% \\
& 1 - 3 GeV & 11.0\% & 12.2\% \\
& 3 - 5 GeV & 4.5\% & 7.0\%  \\
& 5 - 7 GeV & 1.5\% & 5.1\% \\
& 7 - 9 GeV & 0.8\% & 3.6\% \\
& \(>\) 9 GeV & 2.0\% & 60.3\% \\
\hline
\multicolumn{2}{c}{Through-going} & 76.3\% & 10.8\% \\
\hline
\multicolumn{2}{c}{Stop} & 3.9\% & 1.0\% \\
\hline
\end{tabular*}
\end{table}

\begin{figure*}[!htb]
        \centering
        \subfigure[]{
            \includegraphics[width=0.4\hsize]{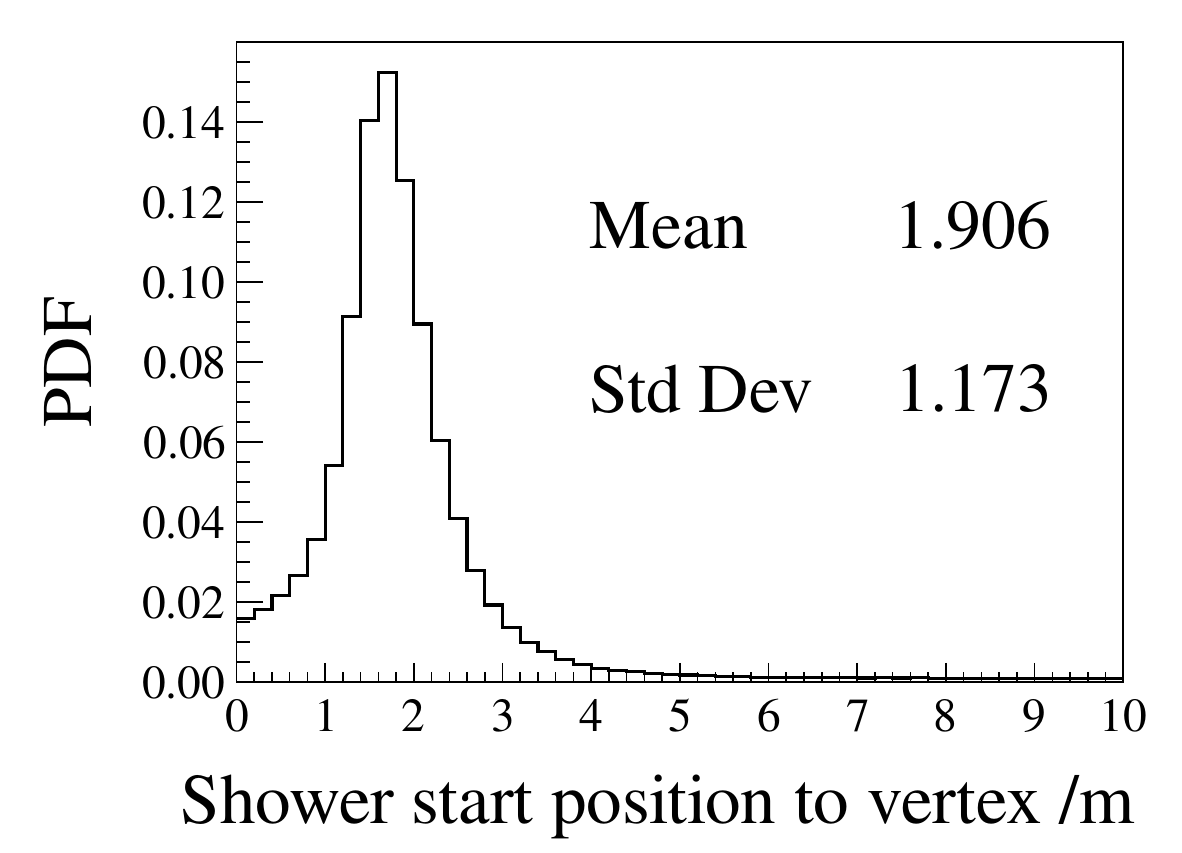}
        	\label{fig:vertex_to_start}
        }
        \quad
        \subfigure[]{
            \includegraphics[width=0.4\hsize]{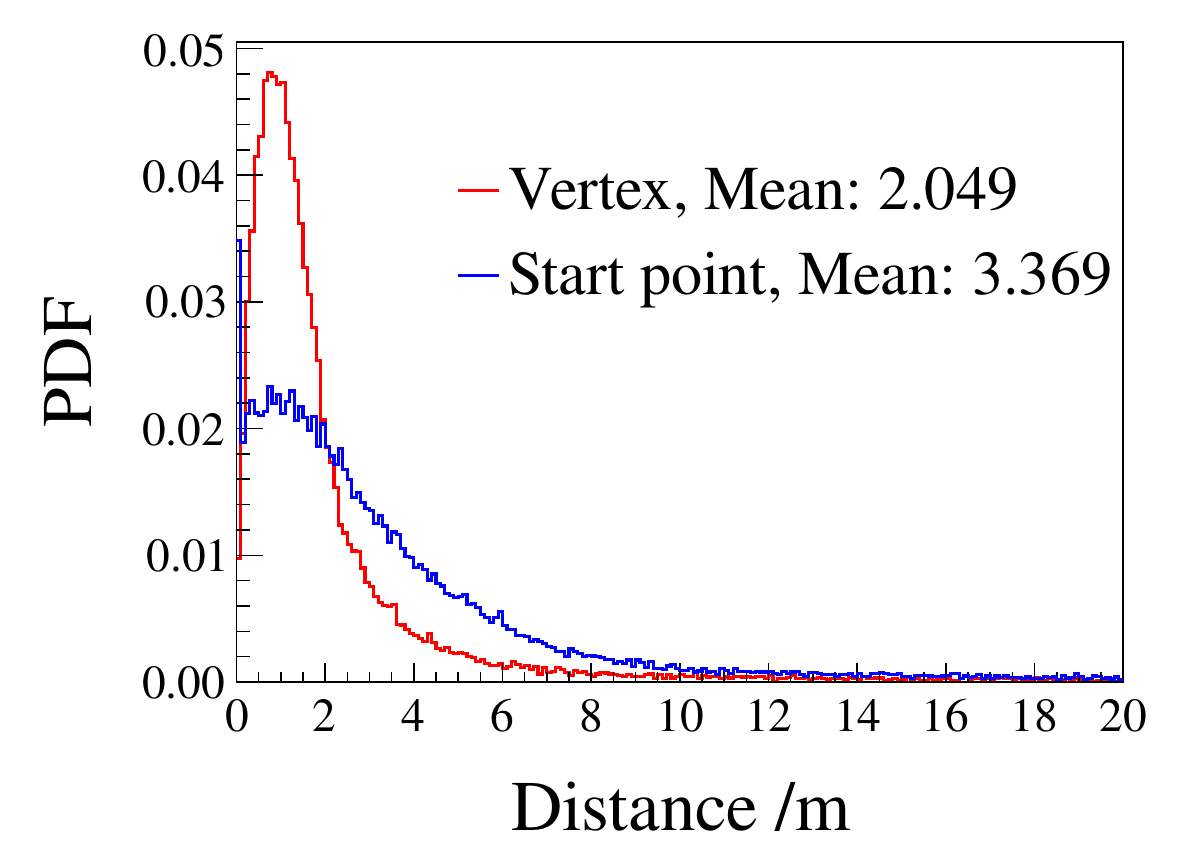}
        	\label{fig:isotopes_to_position}
        }
        \caption{ (a) Distance from shower vertex to shower start position. (b) The distances of the muon-induced isotopes to the shower vertex (red line) and to the initial interaction position of the shower (blue line) }
    	\label{fig:voxelprofile1}
\end{figure*}

Table~\ref{tab:iso} shows the proportion of isotopes produced by different shower energies of muon events and by non-shower muon events. Although the shower muon events occupy the ratio of 19.8\% in all muon events, they can contribute about 88.2\% of the isotopes. Consequently, if the shower vertex can be reconstructed, then the isotope background caused by muons in neutrino analysis can be significantly reduced. As mentioned in Section~\ref{sec:simulation}, the shower vertex discussed in this paper corresponds to the weighted center of the shower energy deposition. From Fig.~\ref{fig:vertex_to_start}, it can be seen that the average distance between the shower vertex and the initial interaction position of the shower (shower start position) is about 1.9~m. Additionally, Fig.~\ref{fig:isotopes_to_position} shows the distances of muon-induced isotopes to the shower vertex and the initial shower interaction position; the proportions of isotopes within 3~m from these two positions are 84$\%$ and 60$\%$, respectively. This also indicates that using the shower vertex as the reference point can more effectively eliminate cosmic isotopes, which is consistent with the conclusion in~\cite{Li-2014sea}.

\subsection {Waveform characteristic caused by shower } 
\label{sec:WF}  

In this section, the waveform characteristics are presented and can subsequently be utilized to reconstruct the shower vertex (Section~\ref{sec:method}).

\begin{figure}[!htb]
\centering
\includegraphics[width=0.7\linewidth]{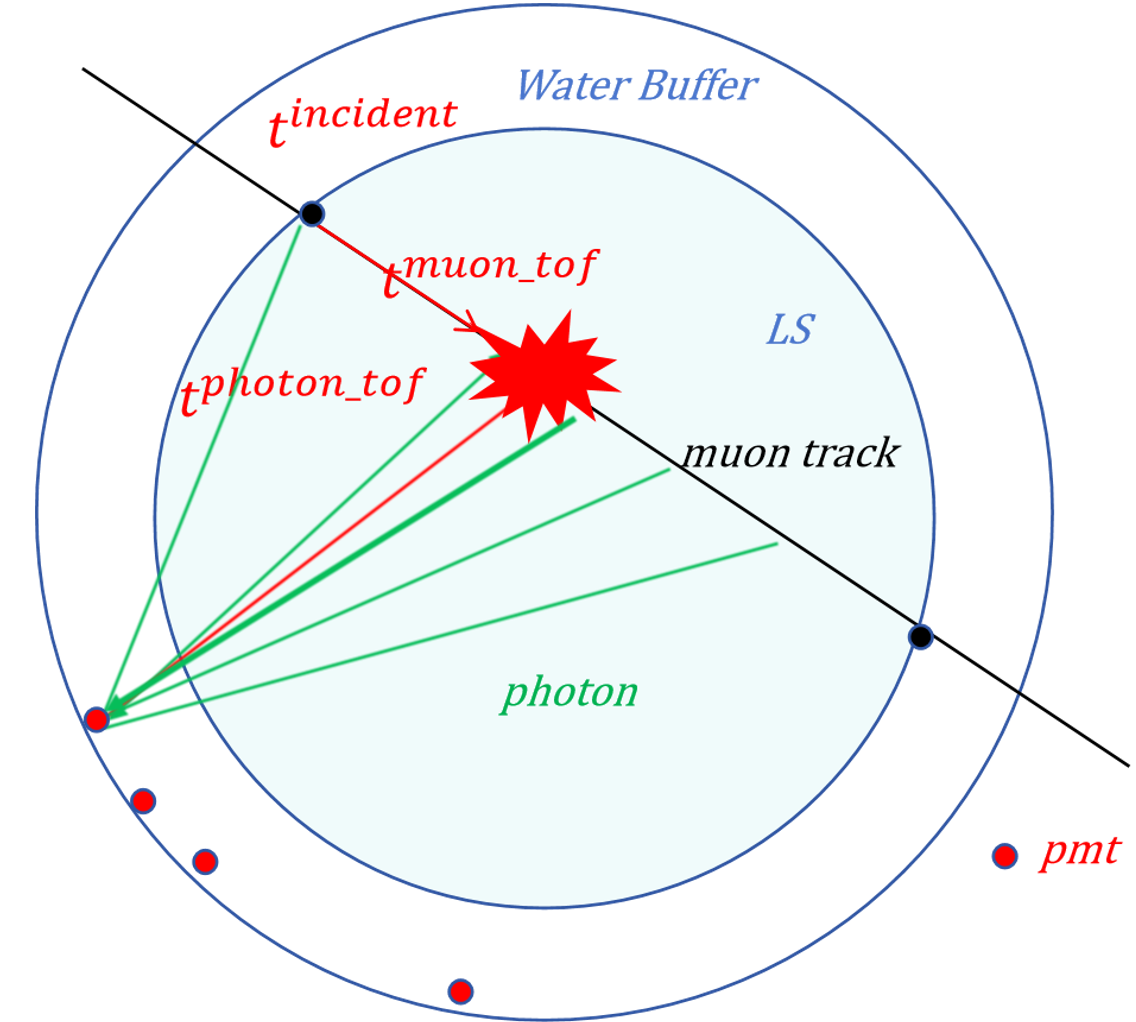}
\caption{\label{fig:predicted_t} Schematic diagram of a muon goes through LS, the light emition and transmits to the PMTs}
\end{figure}

When a muon traverses the LS (Fig.~\ref{fig:predicted_t}), scintillation photons emitted along its path propagate through the medium and subsequently impinge on the PMTs. The combined response of the PMTs and back-end electronics induces measurable waveforms. In the event of a muon-induced shower, substantial energy deposition occurs within the LS. Consequently, a significant flux of photons is generated near the shower. These photons propagate through the LS toward the PMTs. The photon flux from the shower energy deposition introduces distinct peaks into the PMT waveforms~\cite{Huang-2017abb, Tang-2024jfs}, making the waveforms of shower events distinguishable from those of non-shower events. To investigate the waveform features attributable specifically to the shower component, the electronics simulations were performed for both showering and non-showering muons. Within these simulations, the trajectories of shower muon events and non-shower muon events were maintained near-identical by adjusting the primary muon momentum. This design ensured comparable detector responses arising solely from the track (minimum ionization) component. Non-shower muons were selected based on a calculated average dE/dx value approximating the minimum ionization energy loss rate (~2 MeV/cm), derived from the total deposited energy and muon track length. 

\begin{figure*}[!htb]
\centering
\includegraphics[width=0.95\linewidth]{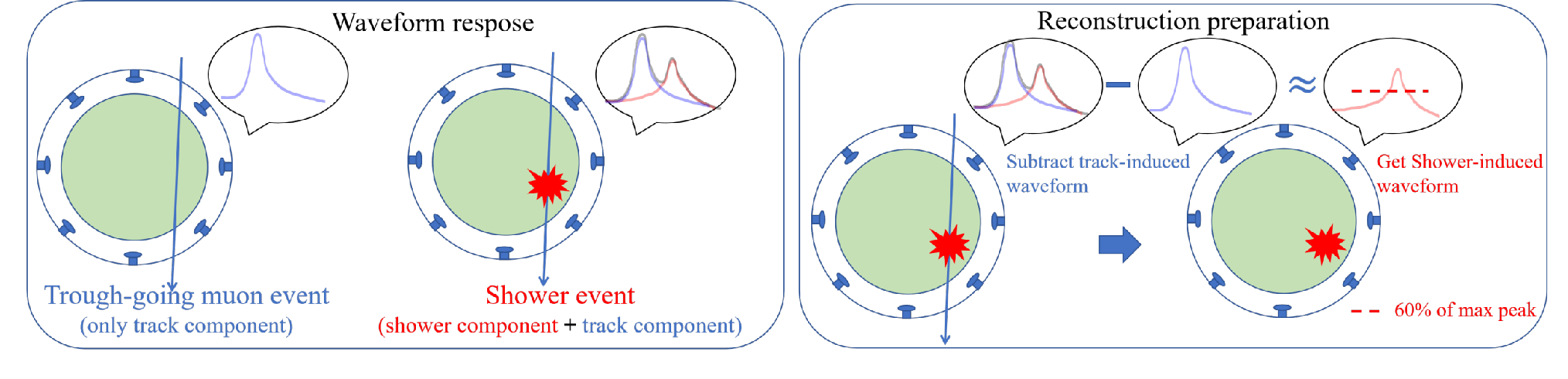}
\caption{\label{fig:subtract_WF} PMT waveforms for a non-shower muon event (blue) and a shower muon event (black). Their trajectories were maintained near-identical by adjusting the primary muon momentum. The waveform of the shower muon event incorporates contributions from both track- and shower-related energy depositions. To enable accurate reconstruction of the shower vertex, the track-induced component is subtracted from the shower waveform by subtracting the waveform of the non-shower muon event, thereby isolating the signal attributable solely to the shower. In the subtracted waveform, the peak-finding threshold (shown as a red dashed line) is set at 60\% of the signal’s maximum amplitude after subtraction}
\end{figure*}

Fig.~\ref{fig:subtract_WF} shows PMT waveforms from a non-shower muon event (blue) and a shower muon event (black). As discussed in Section~\ref{sec:showerCharacter}, the energy deposition of a shower muon comprises a track component and a shower component (Fig.~\ref{fig:longitudinalextension}); both components contribute to its PMT waveform (Fig.~\ref{fig:subtract_WF}). For accurate shower vertex reconstruction, waveform components directly associated with shower energy deposition must be isolated. This necessitates the prior subtraction of waveform contributions arising from the track energy deposition. By subtracting the waveform of a non-shower muon event from that of a shower muon event, interference from the track component can be effectively mitigated during shower vertex reconstruction.

Due to inherent fluctuations in processes such as energy deposition, photon propagation, and photon detection, even identical muon tracks can yield slightly different waveforms on the same PMT. To reliably identify residual waveform components generated by the shower after track subtraction, a selection threshold was implemented. Candidate peaks attributed to the shower component were required to exceed 60\% of the maximum peak height observed in the subtracted waveform.

\section{Methodology } 
\label{sec:method}

\subsection{Principle}
\label{sec:principle}

The peak information after shower and non-shower subtraction (Fig.~\ref{fig:subtract_WF}) will be utilized for reconstructing the shower vertex. The number of peaks in the subtracted waveform indicates the number of shower vertices that can be reconstructed. By comparing the observed peak times with those predicted from the time evolution of the muon-induced photon propagation, a $\chi^{2}$ function is constructed. This function is summed over all selected PMTs to fully exploit the detector information. For the $k^{th}$ peak in the subtracted waveform, the $\chi^{2}$ is expressed as:

\begin{equation}
\label{equ:chi}
\chi^{2} = \sum_{i} \Big(\frac{ T_{i}^{\rm pre}- T_{i}^{\rm obs}}{\sigma_{i}} \Big)^{\rm 2}
\end{equation}
    
    \noindent
where  $T_{i}^{\rm pre}$ is the predicted $k^{th}$ peak time, $T_{i}^{\rm obs}$ is the observed $k^{th}$ peak time, and $\sigma_i$ represents the uncertainty in the predicted peak time for the $i^{th}$ PMT. In this work, we use the TTS of PMT as the dominant contribution to this uncertainty. A more comprehensive uncertainty model, incorporating contributions from track reconstruction and optical model variations, is left for future study. It is assumed that the $\sigma_{i}$ remains constant for a given PMT throughout the analysis. The $k^{th}$ shower vertex parameters are reconstructed by minimizing the $\chi^{2}$ function, thereby combining information from all selected PMTs to achieve an optimal vertex estimate.

\begin{equation}
\label{equ:tpre}
T^{\rm pre} = t^{\rm incident} + t^{\rm muon\_tof} + t^{\rm photon\_tof} + t^{\rm delay} + t^{\rm offset}
\end{equation}
 
Based on the muon track and the photon propagation process shown in Fig.~\ref{fig:predicted_t}, it can be concluded that the $ T^{\rm pre}$ is composed of muon incident time ($ t^{\rm incident}$), time-of-flight(tof) of muon ($ t^{\rm muon\_tof}$), time-of-flight of photons ($ t^{\rm photon\_tof}$), delay time caused by the PMT, cable and readout electronics ($ t^{\rm delay}$) and the offset of the first hit time (FHT) and peak time of the waveform ($\rm t^{\rm offset}$), as indicated in Eq.~\eqref{equ:tpre}.

\begin{figure*}[!htb]
\centering \includegraphics[width=0.8\linewidth]{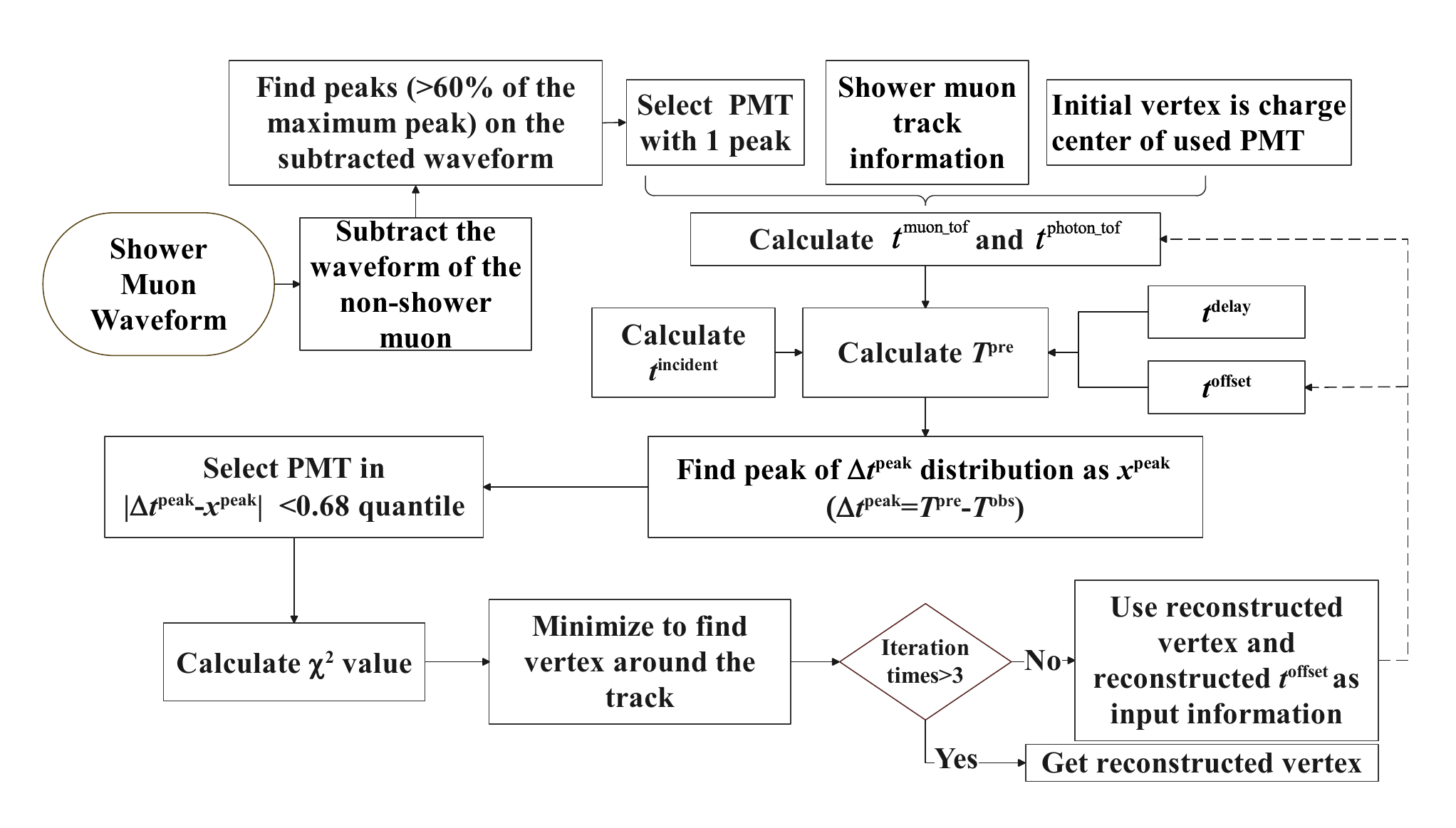}
\caption{\label{fig:workflow} Reconstruction process}
\end{figure*}

The overall reconstruction process is summarized as shown in the Fig.~\ref{fig:workflow}. Following electronics simulation, the PMT waveforms are obtained. The waveforms of the shower muon events need to be subtracted from those of the non-shower muon events. The method for obtaining the waveforms of non-shower muon events will be explained in Section~\ref{sec:noshowerWFmethod}. The TSpectrum tool in the ROOT package~\cite{Brun-1997pa,root-tool} is used to find the peaks over threshold in the subtracted waveform. In the reconstruction of a single shower vertex, PMTs with a single peak after subtraction are selected, and the peak time is $T_{i}^{\rm obs}$ as described in Eq.~\ref{equ:chi}. 

From Fig.~\ref{fig:predicted_t}, it is best to have muon track information as input for calculating the $t^{\rm muon\_tof}$ and the $t^{\rm photon\_tof}$. The track of the shower muon event can be reconstructed using the fastest light method \cite{Genster-2018caz,Zhang-2018kag} or machine learning method \cite{Liu-2021okf}. If muon track is not available, it is necessary to search throughout the entire detector volume during the minimization process of Eq.\ref{equ:chi} to reconstruct the shower vertex. The initial value of the shower vertex will be provided by the charge centroid method. During the minimization process, $t^{\rm muon\_tof}$ changes as the candidate shower vertex iterates to different positions. The propagation of photons in LS~\cite{Zhou-2015gwa, Gao-2013pua, Zhang-2019lrh, Wurm-2010ad, Zhang-2020mqz, Ding-2015sys, Buck-2015jxa, OKeeffe-2011dex,Zhang-2024refractive,Beretta-2025fluorescence} is complicated and includes various optical processes including scintillation, Cherenkov process, absorption and re-emission, Rayleigh scattering, and reflection or refraction at detector boundaries formed with acrylic and water materials. The time-of-flight may not be proportional to propagation distance. In the calculation of the time-of-flight, the effective refractive index $n_{\rm eff}$ is required for a multilayered medium, where it depends on the propagation distance~\cite{Huang-2022zum}. For a certain PMT, $t^{\rm photon\_tof}$ can also be calculated based on the shower vertex position and the effective refractive index. The PMTs densely cover about 78\% surface of the detector, therefore, the position where the earliest photons hit the PMTs approximately corresponds to the muon incident position, and the time of this hit is also approximately the muon incident time. The FHTs of all PMTs are sorted from smallest to largest, and the average of the first five PMTs is taken as the $t^{\rm incident}$. The $t^{\rm incident}$ calculated using the average value is closer to the true incident time, which can avoid the fluctuation when using the minimum FHT as the muon incident time. The readout electronics and signal transmission cables of the PMT can cause a time delay ($t^{\rm delay}$). The $t^{\rm delay}$ can actually be calibrated using a laser calibration source. The $t^{\rm offset}$ is set as a variable to describe the difference between the waveform peak time and the FHT. This initial value of $t^{\rm offset}$ can also be estimated from the waveform and bounded within a specified range during minimization. Based on Eq.~\eqref{equ:tpre}, the expected time ($T^{\rm pre}$) of the peak caused by the shower on the waveform of each PMT can be predicted. 

\begin{figure}[!htb]
\centering
\includegraphics[width=0.9\linewidth]{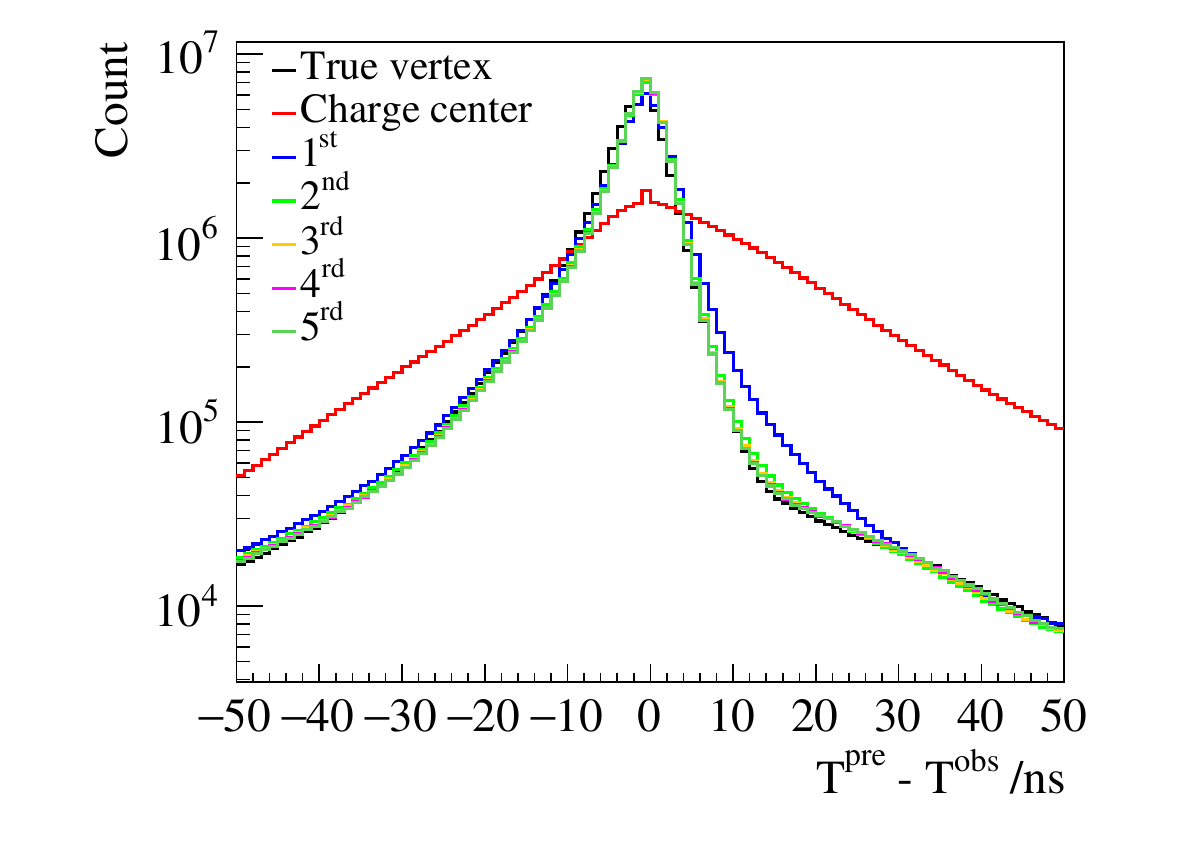}
\caption{\label{fig:del_t}  $T^{\rm pre}$ - $T^{\rm obs}$ distribution with different iteration times during reconstruction}
\end{figure}

The difference between $T^{\rm pre}$ and $T^{\rm obs}$ determined by peak finding in the subtracted waveform, results in a distribution of $\Delta t^{\rm peak}$ ($\Delta t^{\rm peak}$ = $T^{\rm pre}$ - $T^{\rm obs}$), as indicated by the red curve labeled "charge center" in Fig.~\ref{fig:del_t}. Since the value of $\Delta t^{\rm peak}$ can assess the accuracy of the $T^{\rm pre}$, it is also used to select suitable PMTs as input for shower vertex reconstruction. Through in-depth optimization studies, it was determined that PMTs within a region covering 68\% of the distribution around the peak value of $\Delta t^{\rm peak}$ serve as inputs for vertex reconstruction. 

Furthermore, based on the true shower vertex information and assuming the shower is a point source, the expected distribution of true $\Delta t^{\rm peak}$ can be predicted. Comparing the expected $\Delta t^{\rm peak}$ distribution from the reconstructed vertex with the true $\Delta t^{\rm peak}$ distribution can evaluate the accuracy of the reconstructed vertex, as shown in Fig.~\ref{fig:del_t}. The initial $\Delta t^{\rm peak}$ distribution predicted based on the charge center differs significantly from the true $\Delta t^{\rm peak}$ distribution, indicating a large deviation of the charge center reconstructed vertex from the true shower vertex.

After calculating the $\Delta t^{\rm peak}$ distribution based on the charge center and the selected PMTs, the shower vertex can be initially reconstructed using Eq.~\ref{equ:chi}. With this first reconstructed shower vertex, a new expected $\Delta t^{\rm peak}$ distribution can be calculated, represented by the blue "$1^{st}$" curve in Fig.\ref{fig:del_t}. As the Fig.\ref{fig:del_t} shows, the $\Delta t^{\rm peak}$ distribution calculated after the first reconstruction is closer to the true $\Delta t^{\rm peak}$ distribution, demonstrating the effectiveness and correctness of the reconstruction process.

Using the first reconstructed shower vertex and $t^{\rm offset}$ as new inputs for the reconstruction algorithm, with updated $T^{\rm pre}$ and the selected PMTs, a second shower vertex reconstruction based on Eq.~\ref{equ:chi} is achieved after optimization. The second $\Delta t^{\rm peak}$ distribution, compared to the first, more closely aligns with the true $\Delta t^{\rm peak}$ distribution, as seen in Fig.~\ref{fig:del_t}. This process is repeated iteratively. As the iteration times increase, the RMS of $\Delta t^{\rm peak}$ becomes smaller, and the mean of $\Delta t^{\rm peak}$ approaches zero. $\Delta t^{\rm peak}$ distribution tends to be similar to the distribution calculated using the true shower vertex. This indicates that the candidate shower vertex is gradually converging toward the true vertex. After four iterations, the reconstructed shower vertex's expected $\Delta t^{\rm peak}$ distribution closely matches the true $\Delta t^{\rm peak}$ distribution. An in-depth analysis of the reconstruction performance with additional iterations (e.g., a fifth iteration) has been performed and found that the improvements of performance beyond four iterations are marginal—the vertex resolution and $\Delta t^{\rm peak}$ distribution remain essentially unchanged. Therefore, considering the balance between reconstruction accuracy and computational cost, the vertex reconstructed in the fourth iteration is adopted as the final result of the entire reconstruction algorithm.

\subsection{Method for obtaining waveform of non-shower muon}
\label{sec:noshowerWFmethod}

As mentioned in the previous Section~\ref{sec:principle}, the initial momentum of shower muons in the generator information was reduced during detector simulation to obtain the waveform of non-shower muons. This adjustment suppressed shower development while aiming to preserve the track component. The dE/dx distribution can be derived from the track length and the total deposited energy. The peak position of this distribution corresponds to the well-known minimum ionizing energy of a muon, which is about 2~MeV/cm. By selecting samples within one sigma of this distribution, non-shower muon events can be obtained. 

Directly identifying both shower event and non-shower muon event occurring along the exact same physical track in real experimental data is infeasible. However, several methodologies can be employed to derive the requisite waveforms of non-shower muon events. 

One method employs analytical calculation and is used in this study. The expected waveform of the track component in a shower muon event can be derived based on physical models of energy deposition, photon propagation process, and the response of the PMT. When a non-shower muon event passes through the LS, it deposits energy primarily through minimum ionizing energy. For simplification of calculations, the muon's track is divided into small segments of 30~cm each. The energy deposited in each segment is estimated as about 60~MeV. The midpoint of each segment is taken to be the emission point of photons produced by the energy deposition. These photons isotropically propagate through the LS and eventually hit the PMT with a certain probability. The probability of photons hitting the PMT is calculated based on the attenuation length of the LS and the spatial acceptance of the photocathode area of a given PMT. As shown in Eq.~\eqref{equ:lsyield}, where $\frac{dE}{dx}$ is the minimum ionizing energy, $L_{\rm segment}$ is the length of the 30~cm segments, $Y$ is the light yield of the LS, 
$L_{\rm att.}$ is the attenuation length of the LS and $DE_{i}$ is the detection efficiency of the PMT for the hitting photons. The other item of Eq.~\eqref{equ:lsyield} describes the spatial acceptance of the photocathode area of a specific PMT for receiving photons at the position of an emitted scintillation photon. In the Eq.\eqref{equ:lsyield}, R is the radius of the PMT's photocathode surface, d is the distance between the PMT and the position of the emitted scintillation photon, and $\theta_{\rm edep, PMT_{i}}$ is the angle between the line connecting a specific PMT and center of detector and the direction from the position of emitted scintillation photon to this PMT. Thus, based on Eq.\eqref{equ:lsyield}, the number of scintillation photons produced in the j-th segment of the muon track that are converted into photoelectrons when they hit the PMTs, denoted as $N_{j}$, can be calculated.

\begin{equation}
\begin{aligned}
\label{equ:lsyield}
N_{j} = Y \times \frac{dE}{dx} \times L_{\rm segment} \times e^{-\frac{d}{L_{\rm att.}}} \times DE_{i}\\ \times \frac{ R^{2}_{\rm PMT_{i}} \times \cos{\theta_{\rm edep,PMT_{i}}}}{4d^{2}}
\end{aligned}
\end{equation}

After obtaining the number of photoelectrons ($N_{j}$) detected by the PMT, it is necessary to consider the timing information of each photon hitting the PMT and the waveform of the single photoelectron (SPE) before predicting the waveform for each PMT. The timing information of photons hitting the PMT can be predicted based on the scintillation photon emission time profile generated in the simulation and the time-of-flight ($t^{\rm tof}_{i,j}$) of photons from the photon emission position to the PMT, as described in the second item of Eq.~\eqref{equ:wf}. The $f_{\rm scint}(t)$ in Eq.~\eqref{equ:wf} represents the scintillation photon emission time spectrum obtained through simulation~\cite{Zhang-2020mqz}. Each photoelectron needs to be convolved with the transit time distribution and the waveform of the SPE, as shown in Eq.~\eqref{equ:wf}. $TT_{i}$ is the transit time, $TTS_{i}$ is the sigma of the transit time distribution, and $W_{\rm SPE}^{i}$ is the SPE waveform of i-th PMT. The transit time distribution and the SPE waveform of each PMT can be obtained through calibration. By combining Eq.~\eqref{equ:lsyield} and Eq.~\eqref{equ:wf}, the track component waveform ($W_{i}$) of i-th PMT in shower muon events can be predicted. 

\begin{equation}
\label{equ:wf}
W_{i} = \sum_{j}N_{j} \times f_{\rm scint}(t- t^{\rm tof}_{i,j} )  \\
\otimes \ Gaus(TT_{i}, TTS_{i}) \otimes W_{\rm SPE}^{i}
 \end{equation}

\begin{figure*}[!htb]
        \centering
        \subfigure[]{
            \includegraphics[width=0.45\hsize]{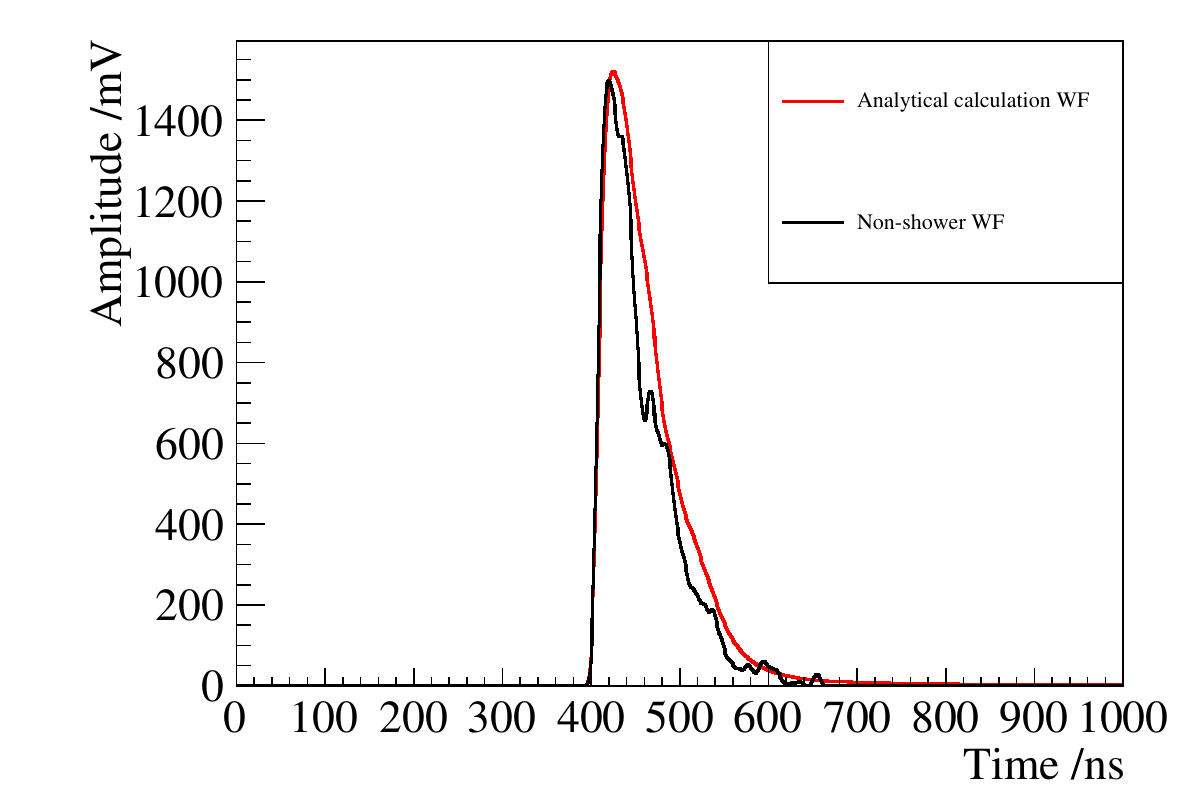}
        	\label{fig:parallel2}
        }
        \quad
        \subfigure[]{
            \includegraphics[width=0.45\hsize]{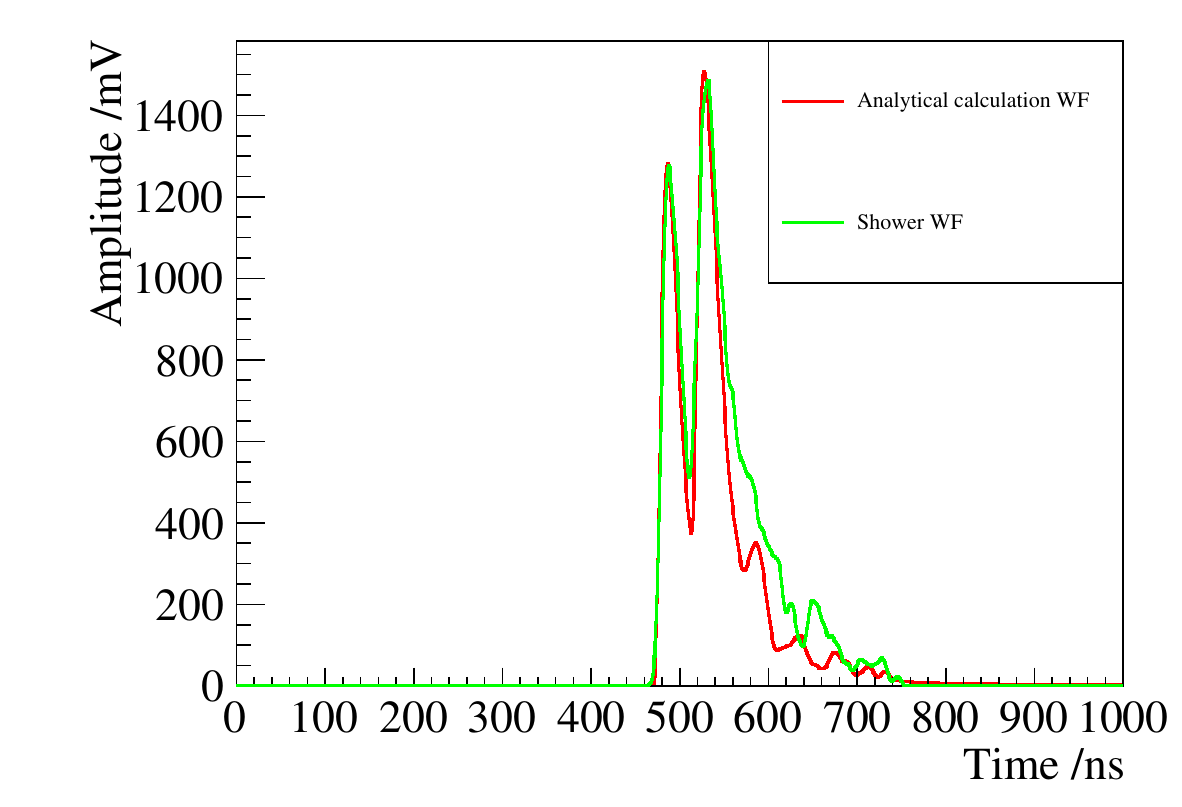}
        	\label{fig:vertical2}
        }
        \caption{ (a) Waveform example for the track component of non-shower muon events. The predicted waveform from analytical calculation is similar to the waveform after electronic simulation. (b) The shower induced waveforms are generated from analytical calculation and after electronics simulation, respectively and they are consistent shape}
    	\label{fig:voxelprofile2}
\end{figure*}

As shown by the red curve in the Fig.~\ref{fig:parallel2}, this is the expected track component waveform of a PMT based on the aforementioned process. The black curve represents the corresponding waveform generated after the detector and electronic simulation. The waveform shape expected from the analytical calculation is generally consistent with the simulated waveform. However, the expected waveform from the analytical calculation may be smoother than the simulated waveform due to the non-uniform energy deposition of the muon along its track and the complexity of photon propagation in the detector. 

If assuming the shower is a point source for depositing energy, the expected waveform of shower component also can be derived based on the Eq.~\eqref{equ:lsyield} and Eq.~\eqref{equ:wf}.  In Fig.~\ref{fig:vertical2}, the red curve represents the expected waveform of shower component for a PMT caused by a shower. The waveform of shower component can be calculated using the shower vertex and the shower energy. The green curve is the waveform of shower component after subtracting the waveform of track component, which has been described in the Fig.~\ref{fig:subtract_WF}. The waveform shapes obtained by both method are basically consistent. This not only demonstrates the effectiveness of the analytical calculation based method for predicting waveform but also confirms that the multi-peak structure of the waveform is indeed caused by the muon shower. Moreover, it shows that subtracting the track component waveform to eliminate the muon's track contribution to the waveform of shower muon events is a feasible approach.

Other methods for obtaining non-shower waveforms are briefly outlined as follows, but further research is needed. For example, the parameter space of reconstructed tracks, such as direction and position, can be segmented into bins aligned with the reconstruction resolution. Muons falling within the same bin are considered to share the similar track. Then, based on the dE/dx values, the non-shower muon data corresponding to those sharing the similar track with the shower muon can be selected, allowing us to obtain the corresponding track component waveforms in real data. In addition, the photon library method, as described in ~\cite{photon}, simulates photon transport via a precomputed lookup table of detection probabilities across the detector volume. This approach can also be used to generate waveforms based on the spatial distribution of energy deposition. Another possibility is to employ machine learning techniques, such as neural networks. They can be employed to learn from a large dataset of simulated samples to generate the expected waveform for a given shower muon track. 

\section{Reconstruction performance}
\label{sec:Performance}

In this section, the performance of the shower vertex reconstruction algorithm is evaluated using simulated samples. As outlined in Section~\ref{sec:method}, accurate reconstruction of the shower vertex requires the prior subtraction of the waveform contribution originating from the muon track. Several subtraction strategies can be employed. To quantify the impact of track accuracy on the performance of the proposed algorithm, we first present results obtained by subtracting the track-component waveforms using the true muon track information from simulation (Section~\ref{subsection:performance_ideal}). We then assess the reconstruction performance when the tracks of showering muon events are smeared to emulate realistic reconstruction uncertainties (Section~\ref{subsection:preformance_track_smearing}).

\begin{figure*}[!htb]
        \centering
        \subfigure[]{
            \includegraphics[width=0.3\hsize]{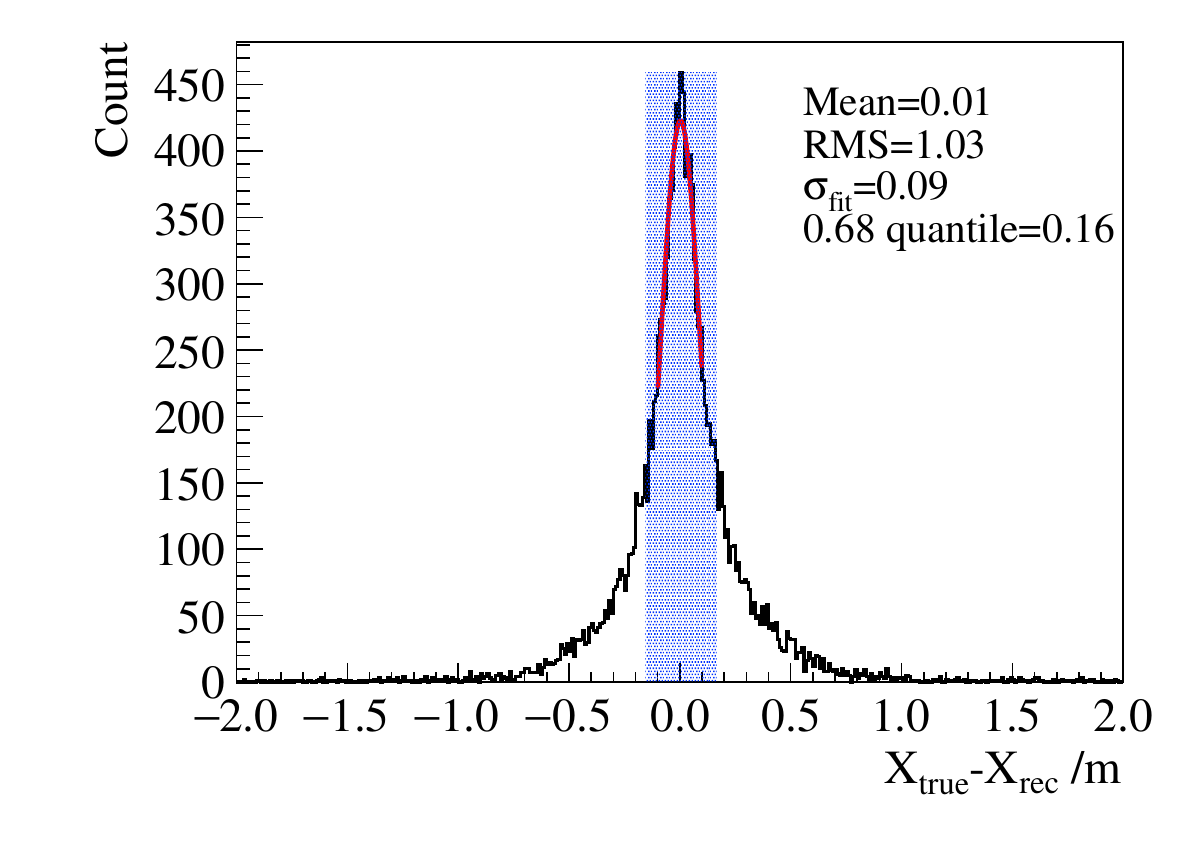}
        	\label{fig:single_rec_difference_distribution_X}
        }
        \subfigure[]{
            \includegraphics[width=0.3\hsize]{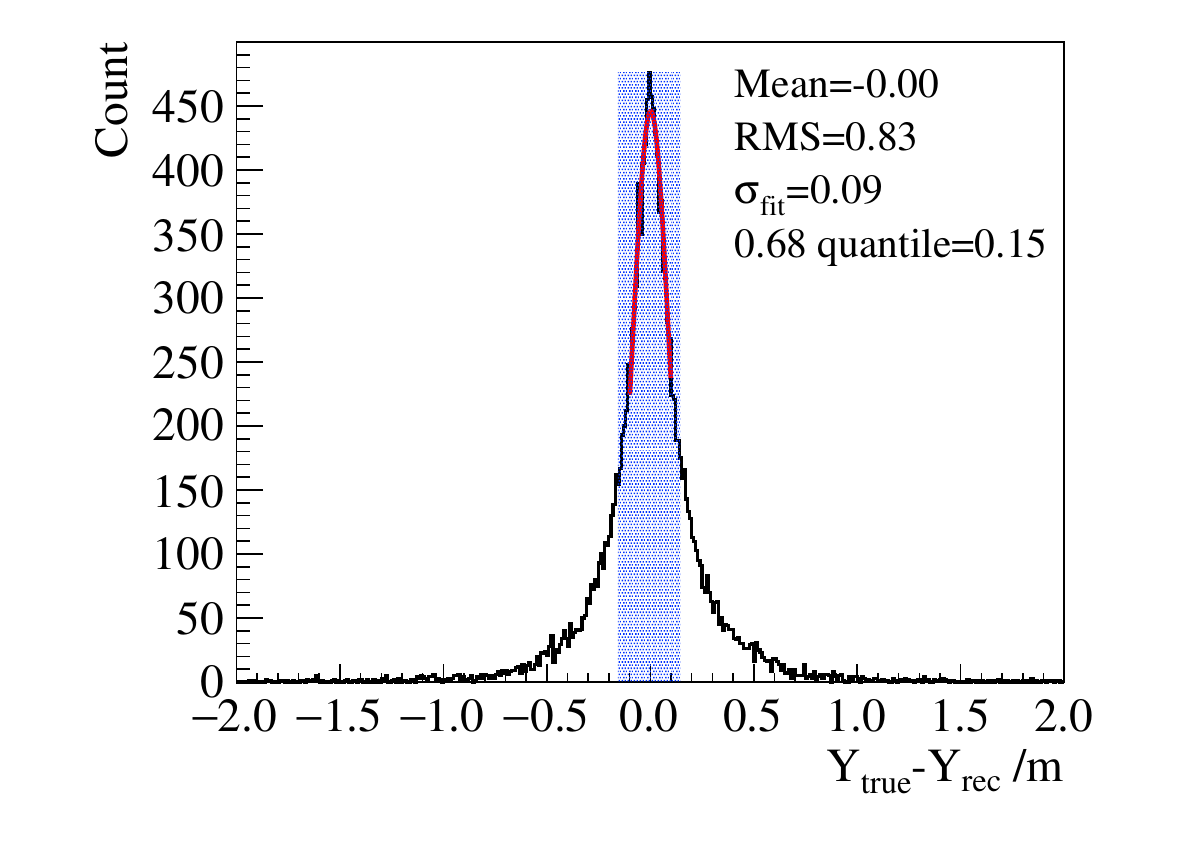}
        	\label{fig:single_rec_difference_distribution_Y}
         }
        \subfigure[]{
            \includegraphics[width=0.3\hsize]{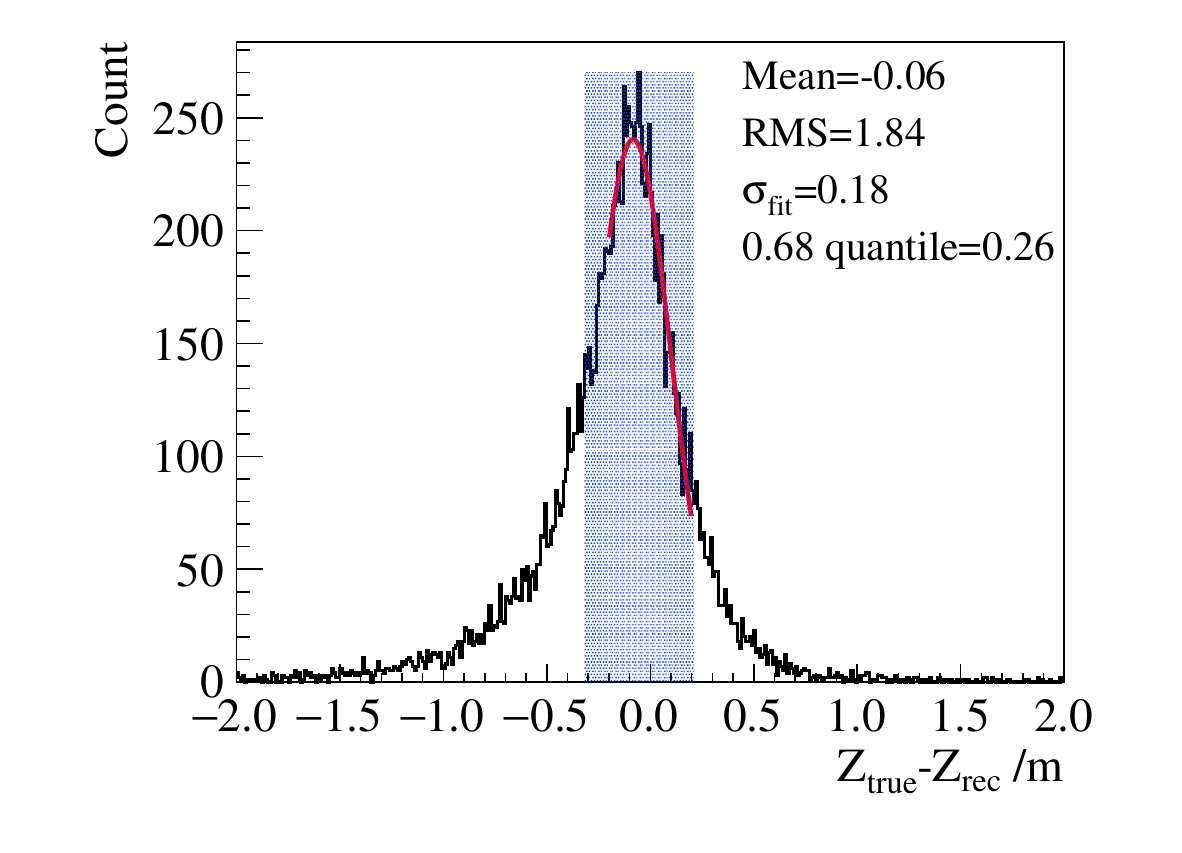}
        	\label{fig:single_rec_difference_distribution_Z}
        }
        \subfigure[]{
            \includegraphics[width=0.3\hsize]{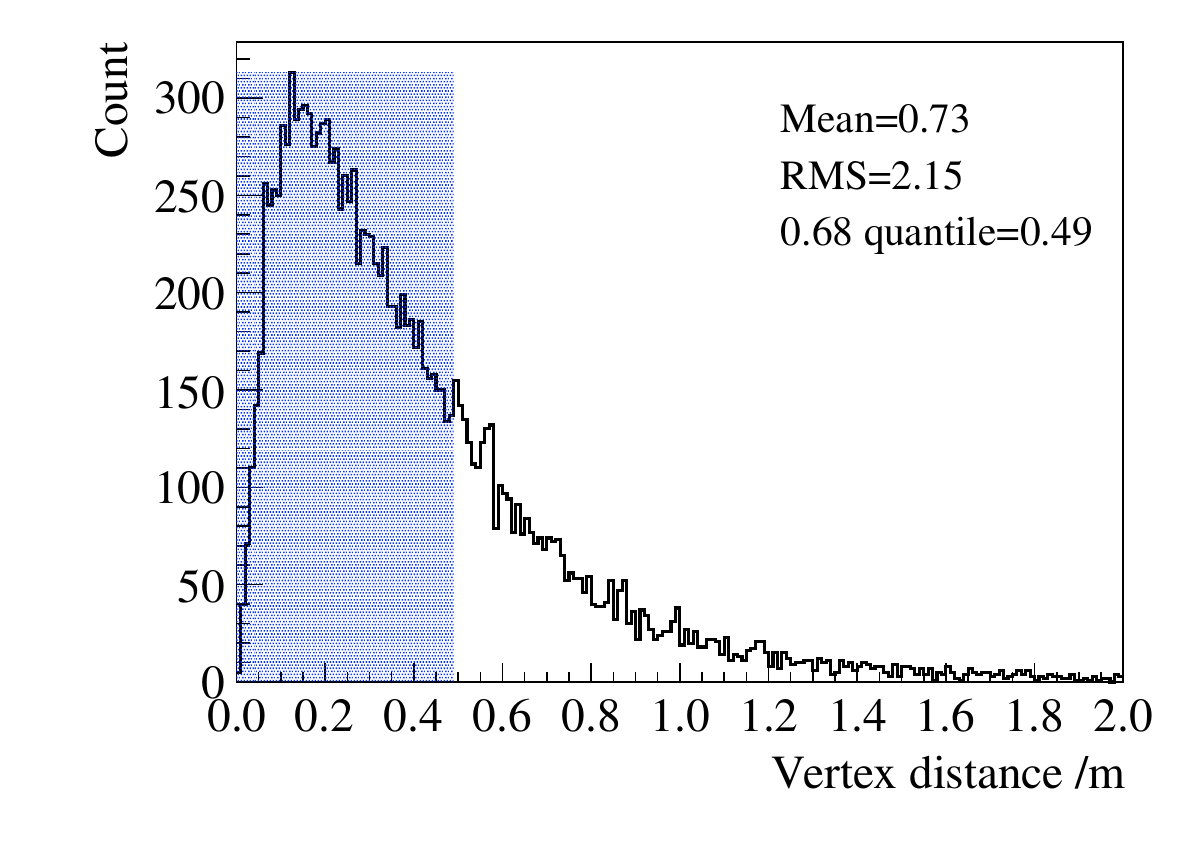}
        	\label{fig:single_rec_difference_distribution_D}
        }  
        \subfigure[]{
            \includegraphics[width=0.3\hsize]{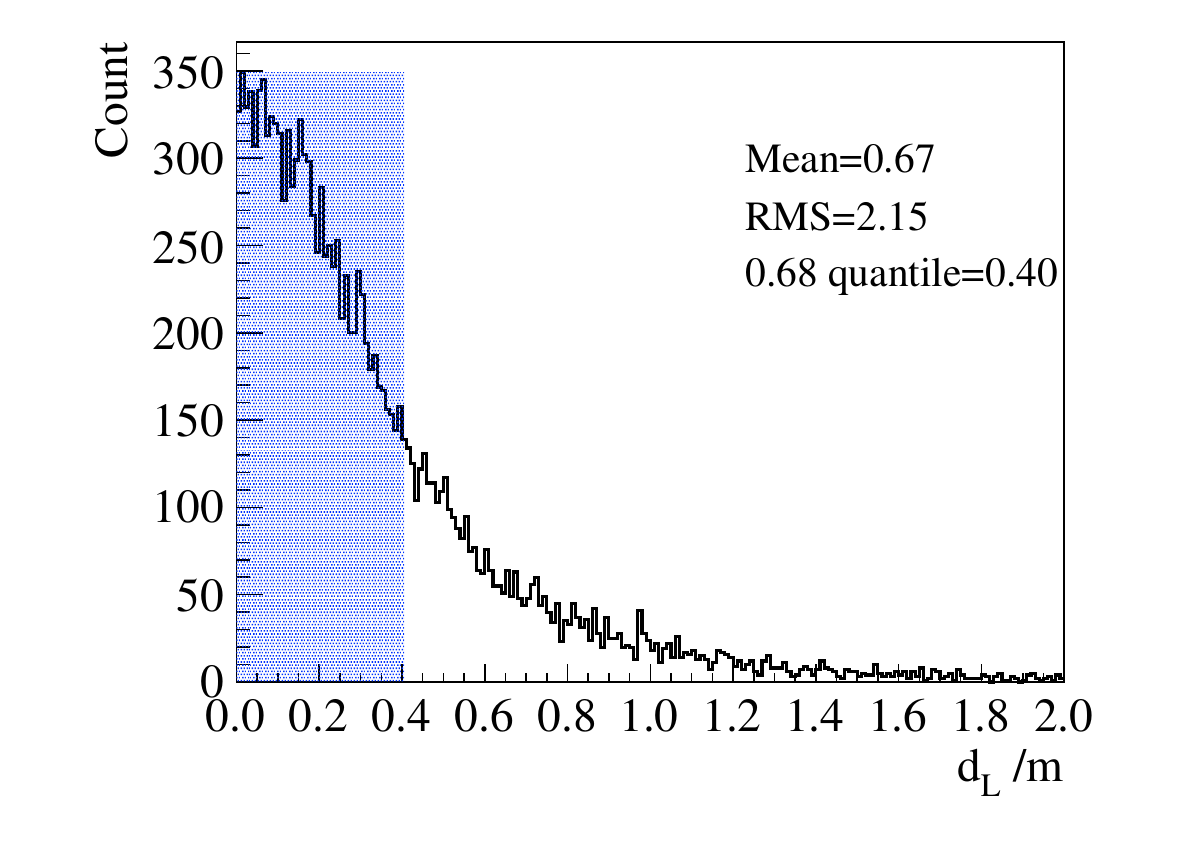}
        	\label{fig:single_rec_difference_distribution_dL}
        }  
        \subfigure[]{
            \includegraphics[width=0.3\hsize]{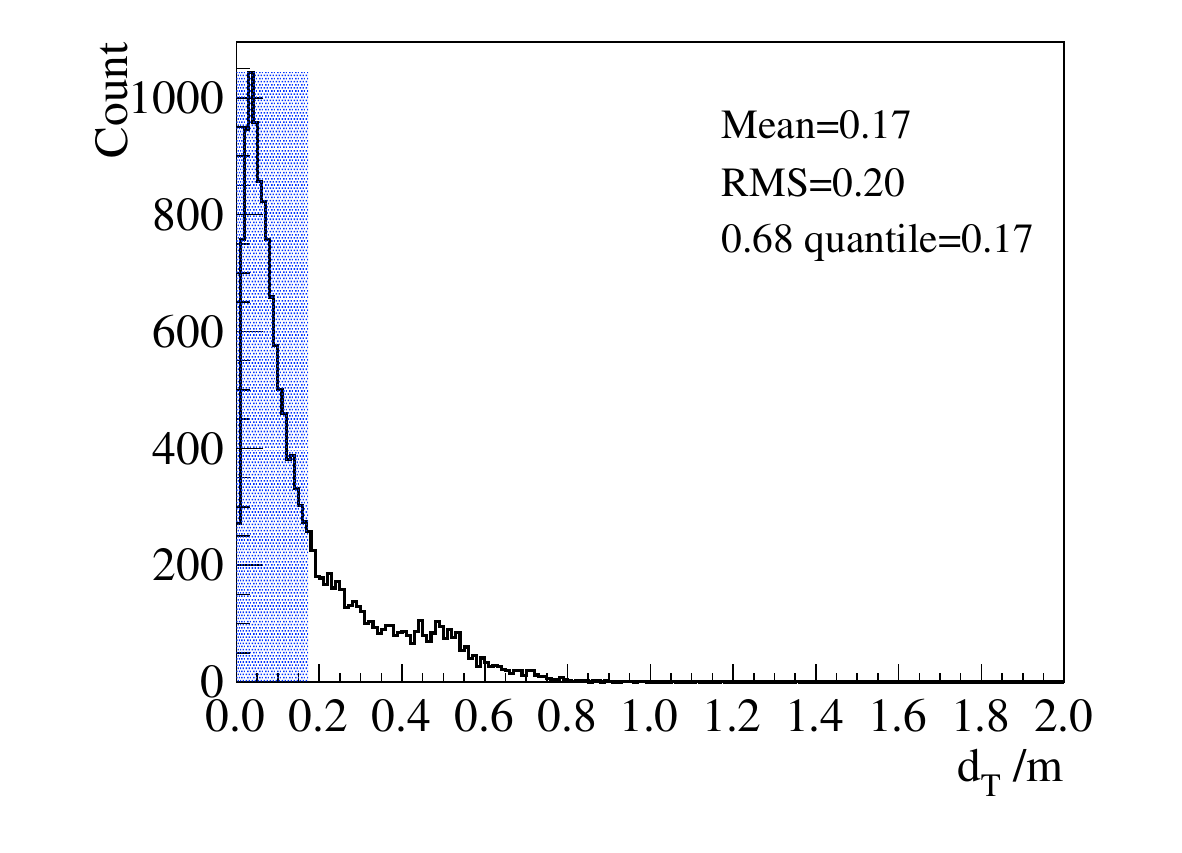}
        	\label{fig:single_rec_difference_distribution_dT}
        }    \caption{\label{fig:single_rec_difference_distribution}  The discrepancies between the true shower vertex and the reconstructed shower vertex. (a), (b), and (c) show the distributions of the reconstruction residuals in the X, Y, and Z coordinates, respectively. (d) shows the distribution of the distance between the reconstructed and true vertices. (e) and (f) show the distributions of the reconstruction residuals projected onto the longitudinal direction (d$_{L}$) and the transverse direction (d$_{T}$), respectively. The red curves represent Gaussian fits to the local area around the peaks in (a), (b), and (c). The shaded regions correspond to the 68\% percentile of each distribution.} 
    	\label{fig:single_rec_difference_distribution}
\end{figure*}

\subsection{Reconstruction performance of single shower}
\label{subsection:performance_ideal}

For evaluating the performance of the shower vertex reconstruction, the Euclidean distance between the true and reconstructed vertices is used firstly, which is a commonly applied metric to quantify spatial distance. Figure~\ref{fig:single_rec_difference_distribution} illustrates the discrepancies between the true and reconstructed shower vertices. For the majority of single-shower events, the reconstructed vertices cluster within a narrow region and exhibit an approximately Gaussian distribution. However, a small subset of events exhibits substantial deviations from the true vertices, indicative of reconstruction failures. The criteria for reconstruction failure and the definition of reconstruction efficiency are detailed in Section~\ref{subsection:Reconstruction_Efficiency}. Gaussian fits were performed near the peak positions of the three spatial component distributions (X, Y, Z). The resulting Gaussian widths ($\sigma$) are 0.09~m (X), 0.09~m (Y), and 0.18~m (Z), significantly smaller than the corresponding Root Mean Square (RMS) values of the distributions (X:~1.03~m, Y:~0.83~m, Z:~1.84~m, vertex distance:~2.15~m). This discrepancy arises primarily from the large deviations of the reconstruction failures. To more accurately characterize the reconstruction performance for the majority of well-reconstructed single-shower events (highlighted by the shaded regions in Fig.~\ref{fig:single_rec_difference_distribution}), the resolution is defined hereafter as the central 68\% percentile of each distribution. Accordingly, the resolutions for the X, Y, and Z components of the reconstructed shower vertex position for all single-shower events are 0.16~m, 0.15~m, and 0.26~m, respectively. For the distribution of the distance between the reconstructed and true shower vertices shown in Fig.~\ref{fig:single_rec_difference_distribution_D}, the resolution defined by the 68\% percentile is 0.49~m. Projecting this distance onto the longitudinal ( d$_{L}$) and transverse ( d$_{T}$) directions relative to the muon track yields resolutions of 0.4~m and 0.17~m, as shown in Fig.~\ref{fig:single_rec_difference_distribution_dL} and Fig.~\ref{fig:single_rec_difference_distribution_dT}, respectively. The larger d$_{L}$ resolution compared to the d$_{T}$ indicates that the shower profile is more extended along the muon direction. This observation, together with the forward-peaking nature of showers, suggests an approximately "egg-like" shower shape.

\begin{figure*}[!htb]
    	\centering
    	\subfigure[]{
        	\includegraphics[width=0.4\hsize]{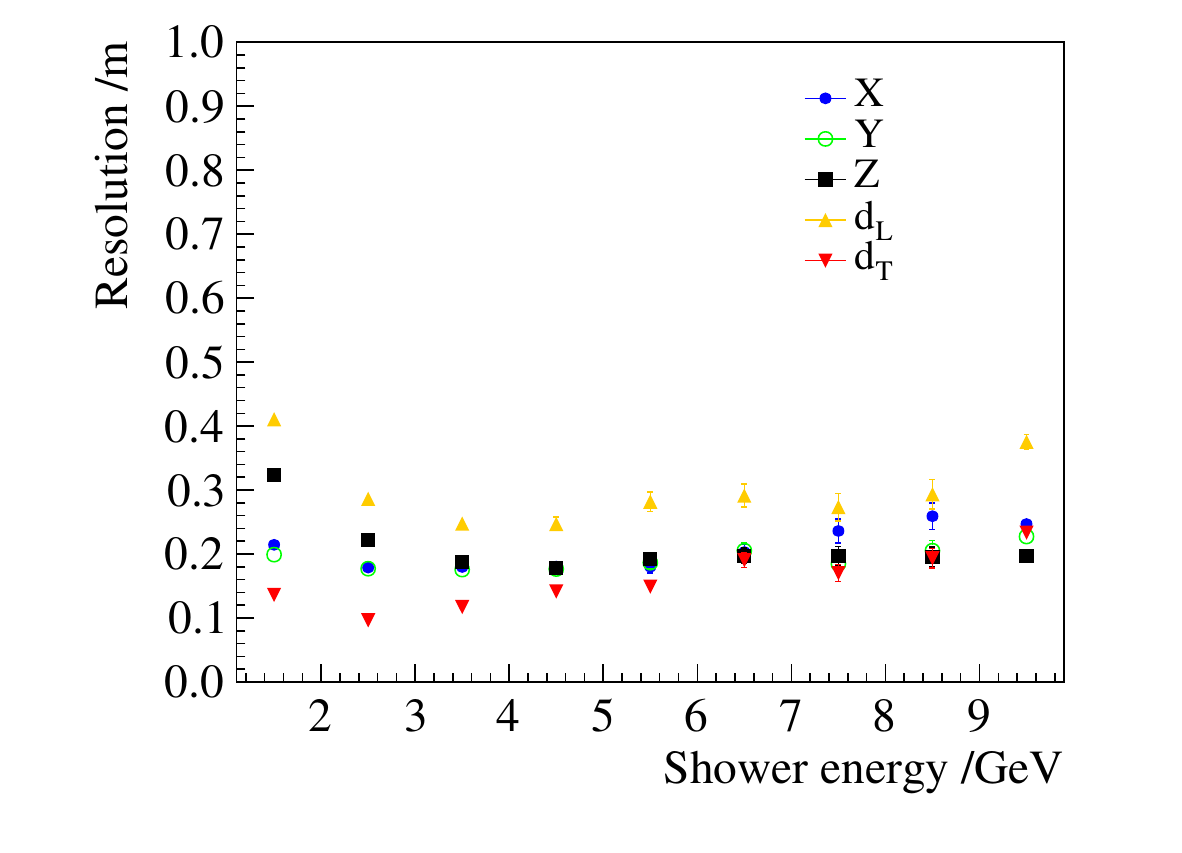}
        	\label{fig:rec-fullsim-E-res}
    	}
    	\subfigure[]{
        	\includegraphics[width=0.4\hsize]{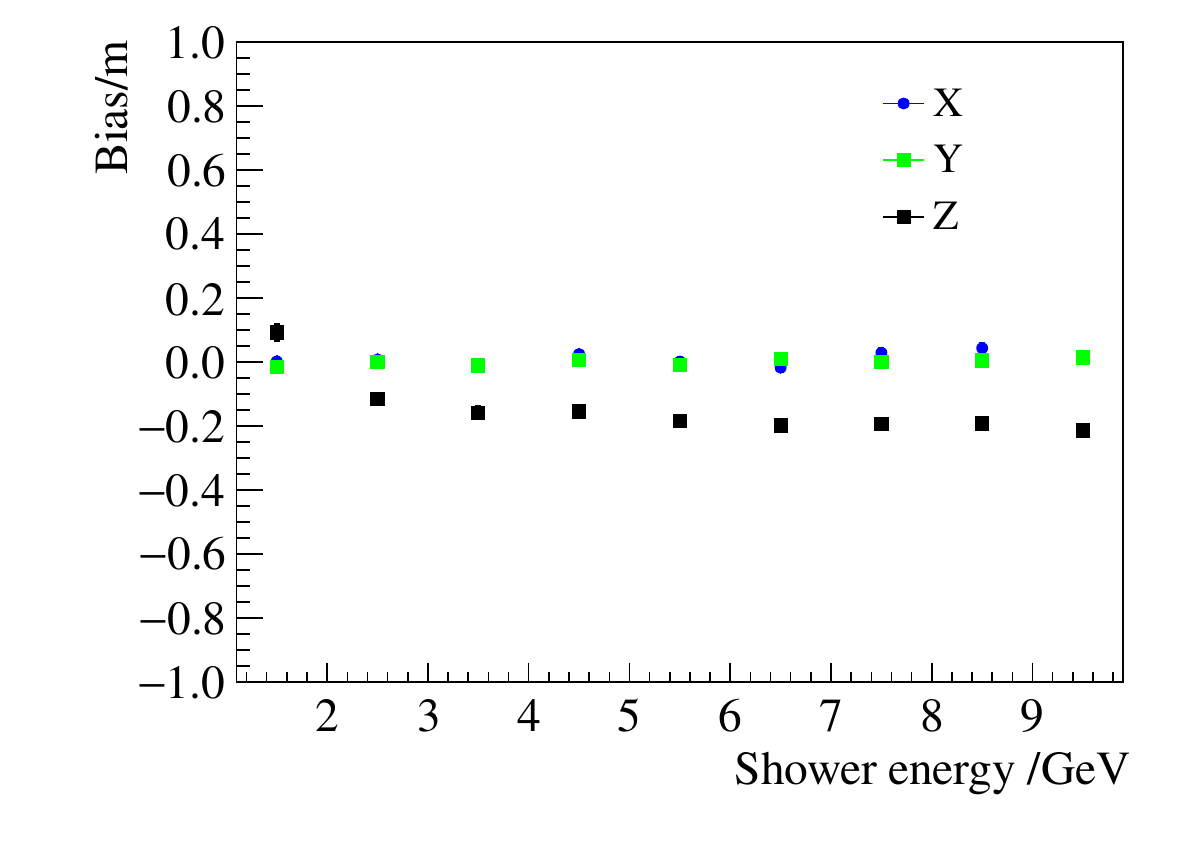}
        	\label{fig:rec-fullsim-E-bias}
    	}
    
    	\caption{(a) The relationship between reconstruction resolution and its shower energy. (b) The relationship between reconstruction bias and its shower. In the above figure, the shower vertex of each individual event in the simulated sample is within a range of 16~m}
    	\label{fig:rec-fullsim-E-performance}
\end{figure*}

Figure~\ref{fig:rec-fullsim-E-performance} presents the dependence of single-shower vertex reconstruction performance on shower energy, specifically examining the resolution (Fig.~\ref{fig:rec-fullsim-E-res}) and bias (Fig.~\ref{fig:rec-fullsim-E-bias}) of the reconstructed vertex position. The spatial resolutions (defined by the 68\% percentile) of the X and Y components remain below 0.2~m, with negligible biases. Moreover, the reconstruction accuracy for the X and Y components shows no significant energy dependence. In comparison, the resolution for the Z component is less precise, though it stays within 0.35~m. The bias in the reconstructed Z position stabilizes at approximately -0.2~m ($Z_{\rm true} - Z_{\rm rec}$) for shower energies above 2~GeV. Detailed analysis attributes the comparatively lower Z reconstruction performance primarily to the forward-peaking nature of muon energy deposition within the shower. This asymmetry in the shower spatial profile along the track direction degrades reconstruction accuracy. Additionally, the higher proportion of muons entering from the detector top (corresponding to smaller zenith angles, $\theta$) further contributes to this effect, as most muons traverse the detector vertically, making the Z direction particularly sensitive to the longitudinal shower asymmetry.

To gain deeper insight into this directional dependence, d$_{L}$ and d$_{T}$ resolution performace shown in Fig.~\ref{fig:rec-fullsim-E-res} also have been examined. Unlike the X,, Y and Z coordinates, which are fixed to the detector, d$_{L}$ and d$_{T}$ are defined relative to the muon direction and thus decouple from the absolute orientation. The  d$_{L}$ resolution improves from about 0.4~m at low energies (below 2~GeV) to approximately 0.25~m at intermediate energies (3–5~GeV), but then slightly degrades to around 0.36~m at higher energies above 9~GeV. This behavior reflects a trade-off: at low energies, the shower-induced waveform features are less pronounced, making it harder to isolate the shower component; at very high energies, the shower develops a larger spatial profile, increasing the effective size of the energy deposition region and thus limiting the precision of its centroid localization. The  d$_{T}$ resolution remains relatively stable around 0.15~m  across most of the energy range, with only a mild degradation at the highest energies due to increased lateral spread. The consistently better  d$_{T}$ resolution compared to  d$_{L}$ confirms the intrinsic elongation of the shower along the muon direction.

\begin{figure*}[!htb]
    	\centering
        	\includegraphics[width=0.4\hsize]{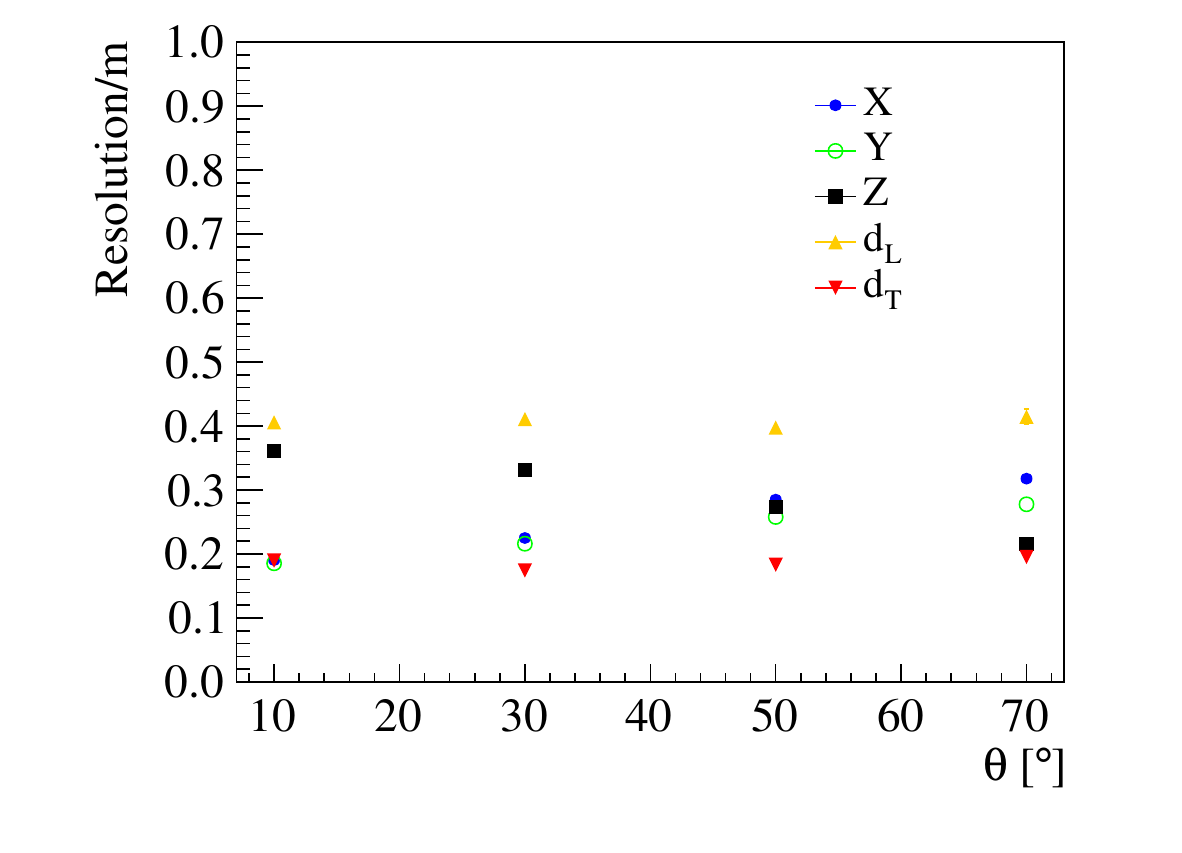}
 \caption{The relationship between the resolution of the shower reconstruction vertex and the incident angle ($\theta$) of the muon}
    	\label{fig:rec_theta}
\end{figure*}

\begin{figure*}[!htb]
    	\centering
    	\subfigure[]{
        	\includegraphics[width=0.4\hsize]{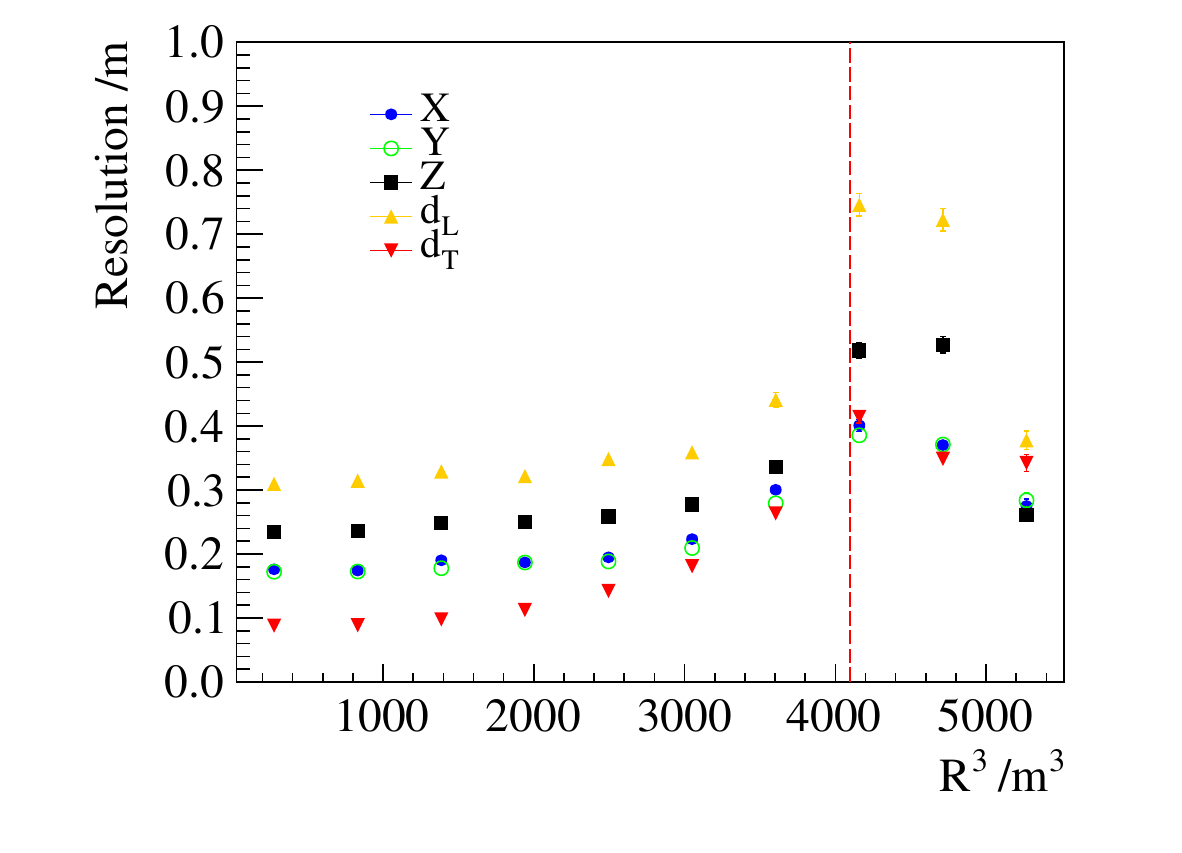}
        	\label{fig:rec_fullsim_R_res}
    	}
    	\subfigure[]{
        	\includegraphics[width=0.4\hsize]{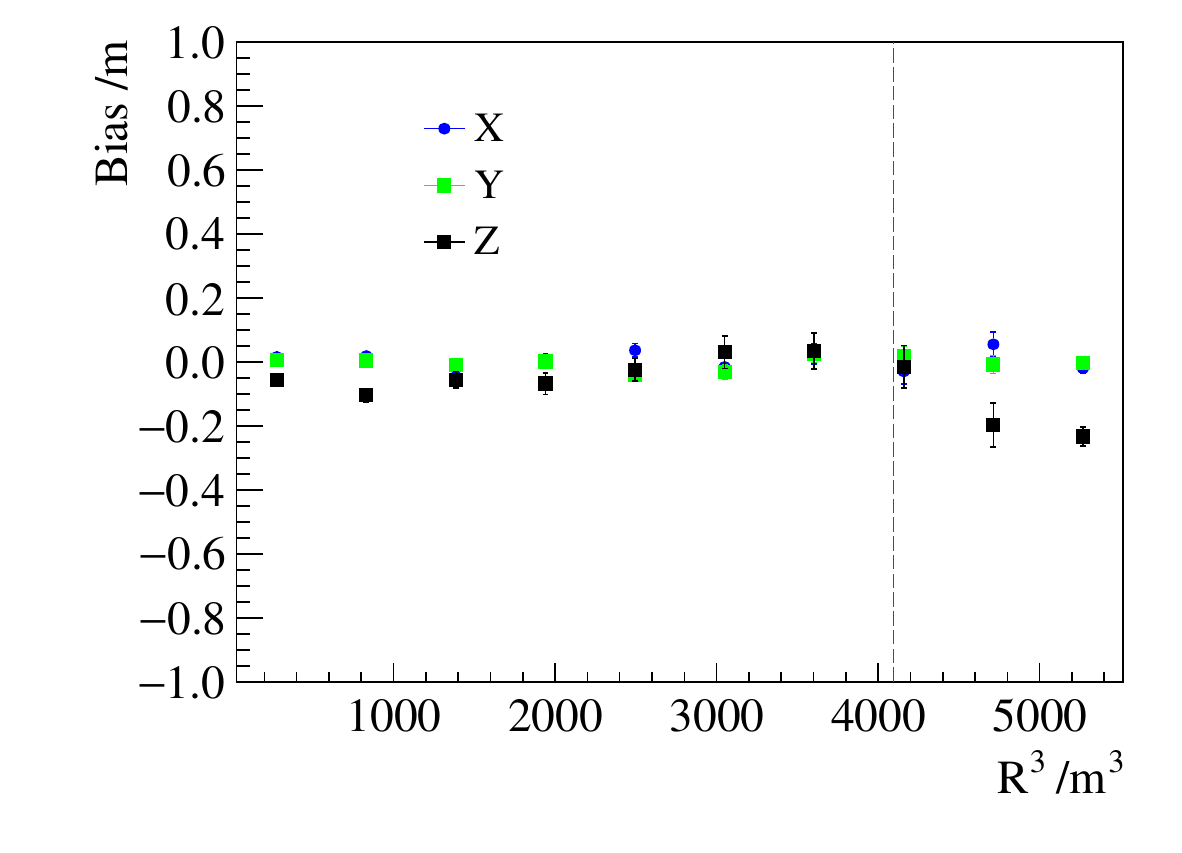}
        	\label{fig:rec_fullsim_R_bias}
    	}
    
    	\caption{(a) The relationship between reconstruction resolution and the vertex position. (b) The relationship between reconstruction bias and the vertex position. On each plot, red vertical dotted lines correspond to $R=16$~m}
    	\label{fig:rec_fullsim_position-VS-performance}
\end{figure*}

The impact of the shower's forward-peaking nature is further illustrated in Fig.~\ref{fig:rec_theta}, which shows the resolution as a function of the muon incident angle $\theta$. The figure shows that the Z-direction resolution of the reconstructed shower vertex improves with increasing muon incident angle ($\theta$). This improvement stems from the forward-peaking of shower energy deposition: as $\theta$ increases, the direction of peak energy deposition shifts increasingly towards the X and Y directions. Consequently, the resolutions along the X and Y axes degrade correspondingly with higher $\theta$. Notably, for $\theta > 50^\circ$, the X resolution deteriorates more rapidly than the Y resolution. Detailed investigations suggest this asymmetry is linked to the non-uniform $\phi$-angle distribution of incident muons, primarily modulated by the mountain overburden. This non-uniformity implies that at larger $\theta$ angles, a higher proportion of shower muons exhibit a more pronounced forward-peak component aligned with the X-direction. As a result, the vertex reconstruction performance (resolution) in X is slightly inferior to that in Y for $\theta > 50^\circ$.

In contrast, the  d$_{L}$ and  d$_{T}$ resolutions exhibit no significant variation with $\theta$. This is expected, as  d$_{L}$ and  d$_{T}$ are defined relative to the muon track direction and thus decouple from the absolute orientation of the detector. Their stability across $\theta$ confirms that the observed variations in the X, Y, and Z resolutions are solely due to the projection of the elongated shower shape onto the fixed detector coordinates. Together, the X, Y, Z and  d$_{L}$/ d$_{T}$ metrics provide complementary views of the reconstruction performance: XYZ is directly applicable to veto strategies in detector coordinates, while  d$_{L}$/ d$_{T}$ reveals the underlying physics of shower development and validates the interpretation of the directional effects.

Figure~\ref{fig:rec_fullsim_position-VS-performance} illustrates the spatial dependence of single-shower vertex reconstruction performance. Reconstruction accuracy remains relatively stable for events occurring within 15~m of the detector center ($R < 15$~m, that is, $R^3 < 3375$~m$^3$), with Z-direction resolution maintained below $\sim$0.3~m and X/Y resolutions below 0.2~m. Corresponding reconstruction biases for all spatial components remain under 0.1~m. Near the detector periphery ($R > 16$~m, that is, $R^3 > 4096$~m$^3$), performance degrades due to muon shower energy leakage and boundary-induced total reflection effects. The distributions of d$_{L}$ and d$_{T}$ exhibit similar spatial dependence to the X, Y and Z directions. Within 15~m of the detector center, the d$_{L}$ resolution remains relatively stable at approximately 0.3~m, but degrades as the vertex approaches the detector periphery. The d$_{T}$ resolution is more sensitive to the vertex position: it remains stable around 0.1~m for vertices within 13~m, but begins to deteriorate beyond this region. This suggests that the transverse reconstruction performance is more susceptible to boundary effects and energy leakage compared to the longitudinal direction.

\begin{figure*}[!htb]
    	\centering
          \subfigure[The relationship between the distance from the reconstructed vertex to the true vertex at different shower energies.]{
        	\includegraphics[width=0.4\hsize]{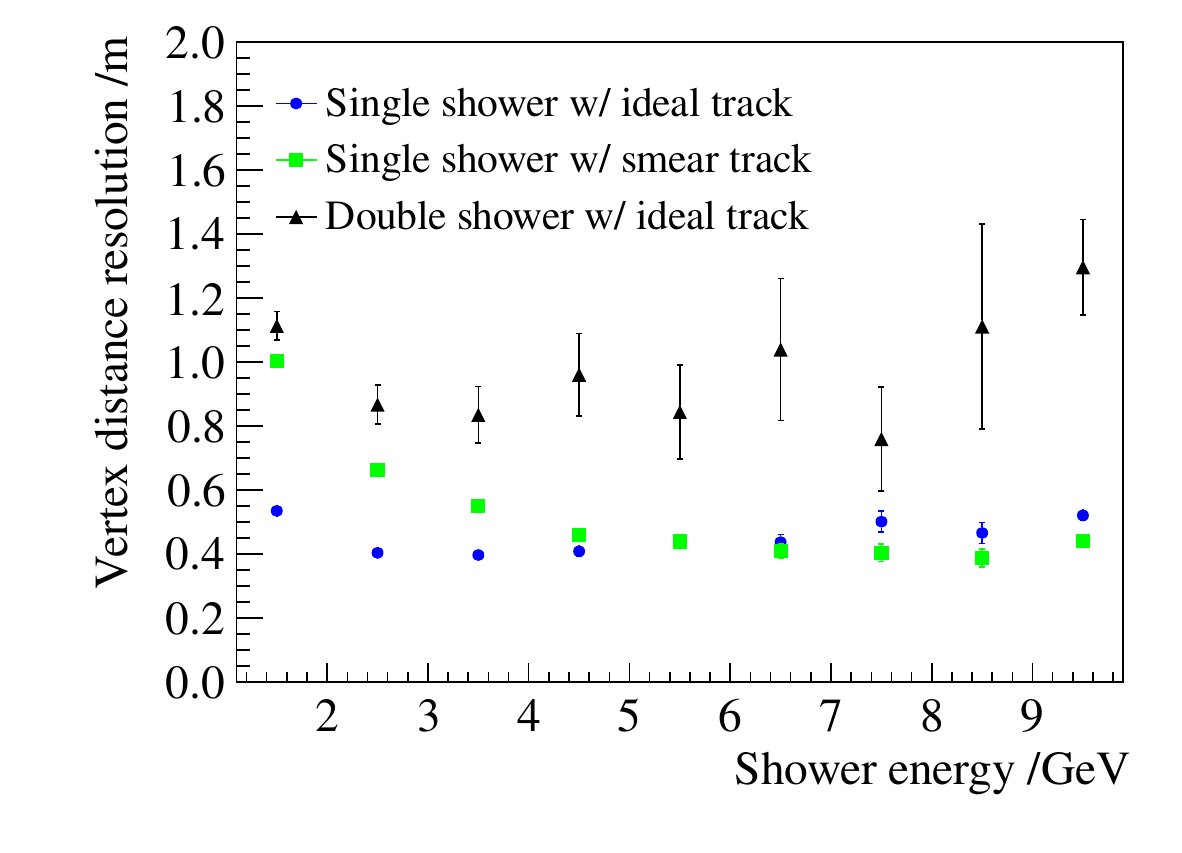}
        	\label{fig:rec_compare_a}
    	}
     \subfigure[The relationship between the distance from the reconstructed vertex to the true vertex at different positions.]{
        	\includegraphics[width=0.4\hsize]{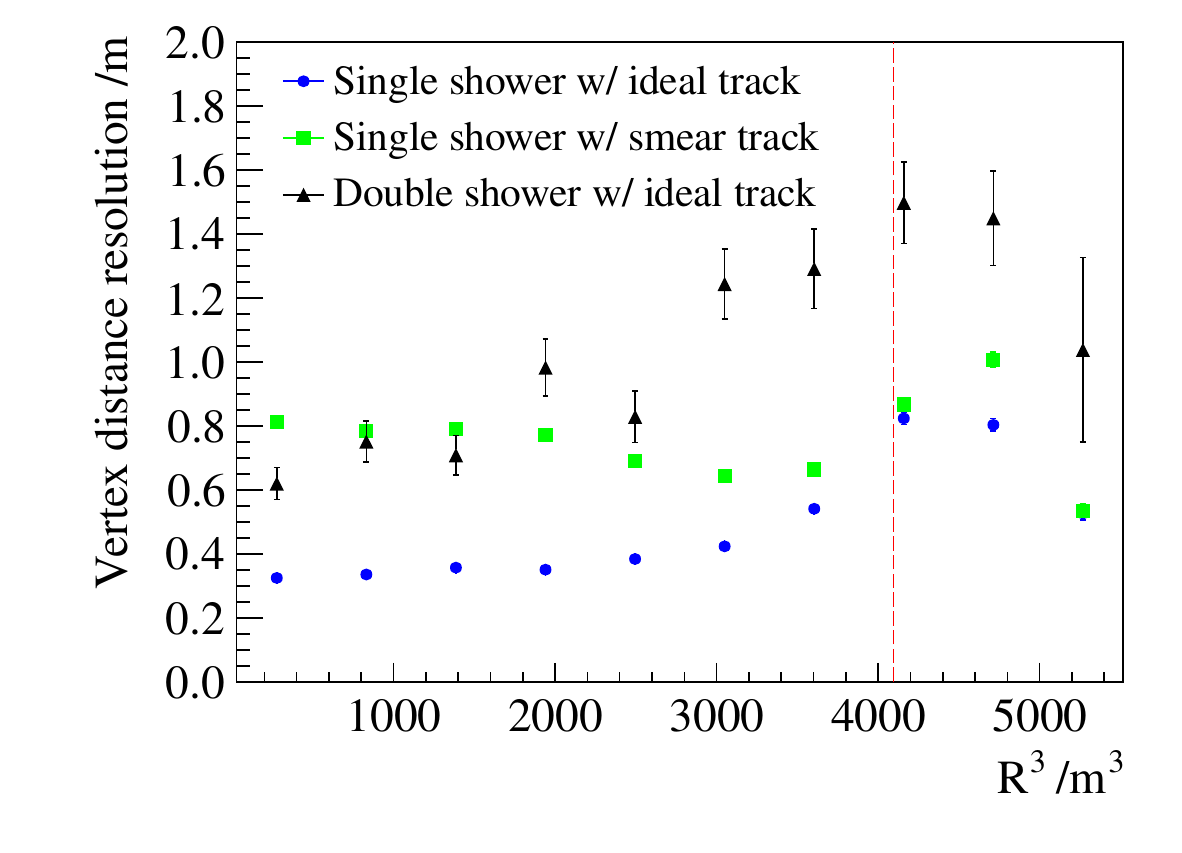}
        	\label{fig:rec_compare_b}
    	}
 \caption{Reconstruction performance of the shower vertex in several scenarios. On the right plot, red vertical dotted lines correspond to $R=16$~m}
    	\label{fig:rec_compare}
\end{figure*}

The blue dots in Fig.~\ref{fig:rec_compare} represent the distribution of the distance between the reconstructed vertex and the true vertex in a single shower event. Figure~\ref{fig:rec_compare_a} and Fig.~\ref{fig:rec_compare_b} respectively show the relationship between this distance and both the shower energy and the vertex position. It can be seen that most of the distances can be controlled within 0.6~m, while they deteriorate to approximately 0.8~m only at positions near the edge of the detector ($R$ > 16~m). Considering the distance relationship between the muon-induced isotopes and shower vertices (Fig.~\ref{fig:isotopes_to_position}) as described in Section~\ref{sec:showerCharacter}, it can be found that the distance between the reconstructed vertex and true shower vertex is less than the distance between the muon-induced isotopes and true shower vertex. This suggests the shower vertex reconstruction algorithm developed in this paper can help enhance the capability to suppress muon-induced isotopes background. On the other hand, this algorithm have potential for reconstructing double-shower events when selecting PMTs with two peaks to be used during the reconstruction process in the Fig.~\ref{fig:workflow}. The black triangles in Fig.~\ref{fig:rec_compare} depict the reconstruction performance of the first shower vertex in a double-shower event. The overall reconstruction performance of the first shower vertex in a double shower event is around 1~m. Regarding the second shower vertex, it is not shown in the figure and its reconstruction performance is affected by interference from the first shower vertex, leading to instability. Therefore, further analysis and investigation are required in the future.

\subsection{The impact of track reconstruction}
\label{subsection:preformance_track_smearing}

The reconstruction results in Section~\ref{subsection:performance_ideal} assume access to the true muon track when subtracting the track-induced waveform. In real experiments, however, the reconstructed track for a shower muon event can deviate from the true trajectory. We therefore assess the impact of track reconstruction uncertainties on shower vertex reconstruction. Following previous studies~\cite{Yang-2022din,Genster-2018caz,Zhang-2018kag,Liu-2021okf}, we adopt a simplified model in which the entry and exit points of the muon track in a single event are independently smeared with a Gaussian of standard deviation $\sigma=20$~cm. The green squares in Fig.\ref{fig:rec_compare} show the results obtained by subtracting the waveform contributions of the smeared track, where the track-component waveforms are computed analytically using Eq.~\eqref{equ:lsyield} and Eq.~\eqref{equ:wf}. The results indicate that, even with this level of track smearing, the distance between the reconstructed and true shower vertices remains within 1~m.

\subsection{Reconstruction efficiency}
\label{subsection:Reconstruction_Efficiency}

\begin{figure}[!htb]
\centering
\includegraphics[width=0.9\linewidth]{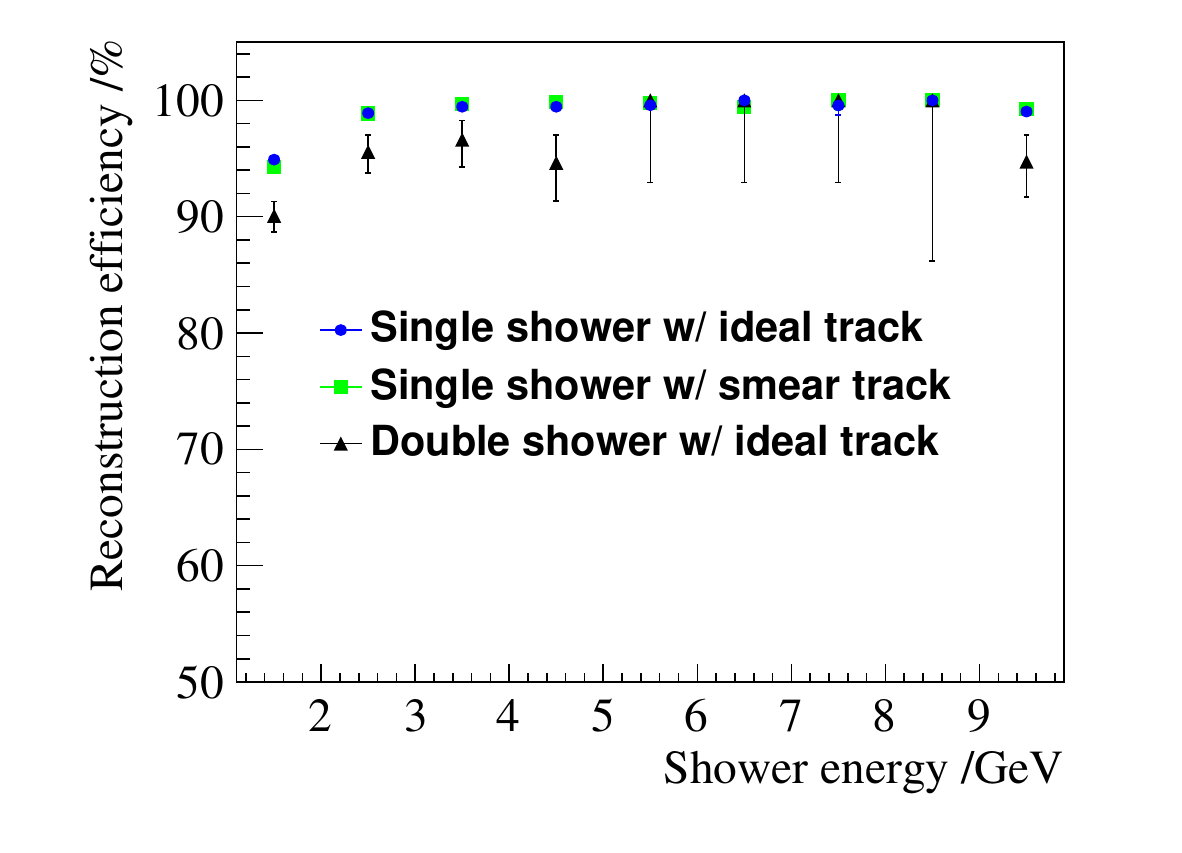}
\caption{\label{fig:rec_eff}  Distribution of reconstruct efficiency vs shower energy}
\end{figure}

As noted in Section~\ref{subsection:performance_ideal}, the RMS of the reconstruction residuals exceeds the Gaussian width ($\sigma$) from the fit (Fig.\ref{fig:single_rec_difference_distribution}), indicating the presence of non-Gaussian tails caused by a small subset of events with large deviations from the true shower vertex. These outliers are attributable to a minor fraction of reconstruction failures. The overall vertex resolution shown in Fig.\ref{fig:rec_compare} remains below 1~m. To establish a clear performance benchmark, we define a successful reconstruction as one for which the distance between the reconstructed and true shower vertices is less than 3m (approximately three times the resolution); events outside this range are classified as failures. Using this criterion, Fig.~\ref{fig:rec_eff} presents the reconstruction efficiency as a function of shower energy. The blue dots and green squares correspond to efficiencies obtained when subtracting waveform contributions computed from the true track and from a smeared track, respectively. The two sets of points largely overlap, and the efficiency approaches 100\% for shower energies above 3~GeV. In particular, for the simulated sample with an initial muon energy of 100~GeV, the single shower reconstruction efficiency reaches 96.7\%. The black triangles in Fig.~\ref{fig:rec_eff} show the efficiency for reconstructing the first vertex in double shower events. As expected from the more challenging event topology, this efficiency is lower than for single shower vertices. Nevertheless, within the 5–9~GeV range, the efficiency for the first vertex in double shower events still approaches 100\%, but statistics in this energy range are limited, the error bars are derived using the Wilson interval method~\cite{Wilson} at a 68\% confidence level. 

In experimental analyses, the $\chi^2 / \mathrm{NDF}$ value serves as a helpful indicator for assessing whether the reconstruction process has been successful. The $\chi^2$ value is the result after minimized the Eq.~\ref{equ:chi} and $\mathrm{NDF}$ is the degrees of freedom used in calculating the $\chi^2$. Using truth information, we partitioned simulated events into two classes: those with a distance between the reconstructed and true vertices greater than 3 m (failures) and those within 3 m (successes). We find that the $\chi^2 / \mathrm{NDF}$ distribution exhibits clear separation: the majority of failures have $\chi^2 / \mathrm{NDF} > 10$, whereas only a small fraction of successful reconstructions exceed this value. This motivates adopting $\chi^2 / \mathrm{NDF} > 10$ as a quality cut to flag unreliable reconstructions in data. Future work could further improve the reconstruction performance by refining both the shower energy-deposition and optical models.

\section{Discussion and conclusion}
\label{sec:conclusion}

In most underground rare-event experiments, muon-induced spallation products constitute a major source of background. We have developed a method for reconstructing muon-induced shower vertices in 20 kiloton-scale LS detectors. The waveform-based algorithm first subtracts the contribution from the through-going muon track; the residual waveforms, which are primarily attributable to the shower component, are then used for vertex reconstruction. The shower vertex is defined as the energy-deposition-weighted centroid, and simulations indicate that approximately 84\% of muon-induced isotopes are produced within 3~m of this point. Given that shower muons ($\sim$19.8\% of all muons) account for $\sim$88.2\% of isotope production, this method enables localized spherical vetoes that suppress the isotope background while preserving signal acceptance and maintaining high physics sensitivity.

To further improve the signal-to-background ratio for isotope identification, an optimized veto strategy has been developed. This strategy utilizes the temporal and spatial correlation to the parent muon, including the muon track information and energy deposition profile. Since decay signals from isotopes can mimic genuine physics signals, an effective veto scheme is essential. Inspired by Ref.~\cite{Abusleme_2021, juno-neutron}, where veto time windows are varied based on the distance from the muon track, we propose applying a similar distance-dependent veto time to the reconstructed shower vertex. For the spherical veto centered on the shower vertex, increasing its radius from 3~m to 5~m is shown to enhance the fraction of vetoed isotopes from 84\% to 92.5\%. The optimal veto radius, however, must be determined in conjunction with the veto time window to balance background rejection against live time. Furthermore, a hybrid veto strategy can be implemented by combining a cylindrical veto volume along the muon track, a spherical veto centered on the shower vertex, and additional spherical vetoes centered on neutron capture locations. The precise veto efficiency and overall performance of this combined approach will require further optimization and study in future work.

For single shower muon events, the algorithm attains spatial resolutions of 0.16~m (X), 0.15~m (Y), and 0.26~m (Z). The resolution for the distance between the reconstructed and true shower vertices is 0.49~m. When projecting this distance onto the longitudinal (d$_L$) and transverse (d$_T$) directions relative to the muon track, the resolutions are 0.4~m and 0.17~m, respectively, reflecting the elongated shape of the shower along the muon direction. The reconstruction efficiency exceeds 96\%. Performance is stable for shower energies above 2~GeV within the central detector region ($R < 16$~m), with modest degradation near the detector boundary. Importantly, when muon-track reconstruction uncertainties are included, the vertex deviation remains within 1~m. The algorithm also shows promise for resolving double shower vertices.

This methodology holds potential for application in other large-scale, low-background experiments that record waveform information, thereby providing a tool for searches targeting neutrinos and dark matter.

\begin{acknowledgements}
We thank the JUNO reconstruction and simulation working group for many helpful discussions. This work was supported by National Key R\&D Program of China No. 2023YFA1606103 and National Natural Science Foundation of China No. 12494574.
\end{acknowledgements}

\end{document}